\documentclass[aps,pra,twocolumn,groupedaddress,floatfix,showpacs,longbibliography]{revtex4-1}
\usepackage{graphicx}
\usepackage{amsmath}
\usepackage{color}
\def\ket#1{|#1\rangle}
\def\bra#1{\langle#1|}

\begin{document}
\title{Spin-Energy Correlation in Degenerate Weakly-Interacting Fermi Gases}

\author{S. Pegahan, J. Kangara, I. Arakelyan and J. E. Thomas}

\affiliation{$^{1}$Department of  Physics, North Carolina State University, Raleigh, NC 27695, USA}

\date{\today}

\begin{abstract}
Weakly interacting Fermi gases exhibit rich collective dynamics in spin-dependent potentials, arising from correlations between spin degrees of freedom and conserved single atom energies, offering broad prospects for simulating many-body quantum systems by engineering energy-space ``lattices," with controlled energy landscapes and site to site interactions. Using quantum degenerate clouds of $^6$Li, confined in a spin-dependent harmonic potential, we measure complex, time-dependent spin-density profiles, varying on length scales much smaller than the cloud size. We show that a one-dimensional mean field model, without additional simplifying approximations, quantitatively predicts the observed fine structure.  We measure the magnetic fields where the scattering lengths vanish for three different hyperfine state mixtures to provide new constraints on the collisional (Feshbach) resonance parameters.
\end{abstract}

\maketitle

\section{Introduction}

Weakly interacting two-component Fermi gases~\cite{DuSpinSeg1}, with tunable, nearly vanishing s-wave scattering lengths $a$, offer a pristine platform for exploring the interplay between spin, motion, and statistics in many-body systems~\cite{KollerReySpinDep}.  In such gases, the collision rate $\propto |a|^2$ is negligible, so that single atom energies are conserved over the evolution time scale set by the mean field frequency $\propto |a|$~\cite{Piechon,MuellerWeaklyInt,DuSpinSeg2,LaloeSpinReph}. Since s-wave scattering in Fermi gases is allowed only for antisymmetric spin states, two-component clouds exhibit an effective exchange interaction, enabling simulations of a variety of spin-lattice models~\cite{KollerReySpinDep}, where the conserved single atom quantum numbers play the role of the lattice sites~\cite{KollerReySpinDep,ThywissenDynPhaseTrans}.  Spin-motion coupling is induced by spin-dependent trapping potentials,  implemented using magnetic field gradients~\cite{Kohl2Dspindynamics,ThywissenTransSpinDyn,ThywissenLeggettRice} or magnetic field curvature~\cite{CornellSpinSeg,LaloeSpinReph,DuSpinSeg1,DuSpinSeg2}. Global spreading of quantum correlations in real space can occur due to the effective long-ranged character of the spin couplings, which is a consequence of the separation of time scales for the fast harmonic oscillation of atoms and slow macroscopic spin density evolution~\cite{LewensteinDynLongRange,KollerReySpinDep}.

The evolution of the spin density in weakly interacting Fermi gases has been described by mean field models employing phase-space representation~\cite{Piechon,MuellerWeaklyInt} and energy representation~\cite{DuSpinSeg2}. The initial implementation of the energy-dependent collective spin-rotation model of Ref.~\cite{DuSpinSeg2} yielded only semi-quantitative agreement with the observed spin-density profiles and the time-dependent amplitude, which were measured at high temperatures, suggesting that the model was incomplete. Recently,  Koller et al.,~\cite{KollerReySpinDep} have devised a new description in terms of Dicke collective spin states, exploiting  conservation of the total spin vector for the exchange interaction. This picture suggests that the observed variation of the spin-wave amplitude with time arises from a thermal average of Dicke gaps~\cite{KollerReySpinDep}, but comparison with the measured spin density profiles has been only qualitative~\cite{KollerThesis}.

\begin{figure*}[htb]
\begin{center}\
\includegraphics[width=4.5in]{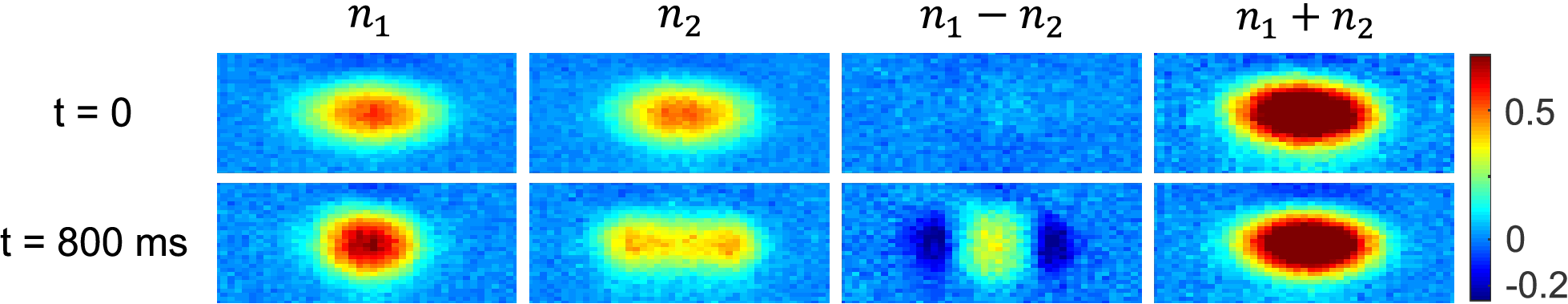}
\end{center}
\caption{Spin-energy correlation  produces spin segregation in a degenerate Fermi gas with an s-wave scattering length of 5.2 bohr. The palettes are $50\times950\,\mu$m. Left to right:  $n_1$, $n_2$, $n_1-n_2$, and $n_1+n_2$ in units of $(n_1+n_2)_{max}$ at $t=0$ (upper) and $t=800$ ms (lower) after coherent excitation of a $|1\rangle-|2\rangle$ superposition state. Note that $n_1-n_2$ evolves in time while $n_1+n_2$ remains constant, due to single particle energy conservation.
\label{fig:Clouds}}
\end{figure*}

We report measurements of time-dependent spin-density profiles for coherently prepared two-state Fermi gases of $^6$Li, confined in a spin-dependent harmonic potential, providing a precise quantitative test of the underlying energy-space spin-lattice model and energy-dependent long-range couplings.  We employ quantum degenerate samples to minimize energy shifts of the scattering length that become significant at higher temperatures. This enables precise comparison of predictions with measured spin-density profiles, which vary from relatively smooth to exhibiting complex structure  over short length scales.   We find that our collective spin-rotation model~\cite{DuSpinSeg2}, extended to degenerate samples, and implemented without additional simplifying approximations~\cite{SupportOnline}, quantitatively predicts the observed spin density profiles. At high temperatures and small scattering lengths $a<1$  bohr, we observe additional new features in the spin-density profiles, which we explain by including the energy dependence of the scattering length in our model.

Using this new model, we determine the zero crossings and magnetic field tuning rates for the s-wave scattering lengths of the three lowest hyperfine states of $^6$Li. Comparing data at high and low temperatures determines the temperature shift of the zero crossings. These measurements provide new constraints on the $^6$Li$_2$ molecular potentials that determine the precise shapes of the Feshbach resonances~\cite{BartensteinFeshbach,JochimPreciseFeshbach}, which have been widely used in studies of strongly interacting Fermi gases~\cite{OHaraScience,BlochReview}. At resonance, where the gas is unitary, the thermodynamic and hydrodynamic properties are universal, depending only on the density and temperature~\cite{HoUniversalThermo}. The most precise measurements of the universal thermodynamic properties~\cite{KuThermo} and of the universal hydrodynamic properties~\cite{JosephAnomalous} rely on the the precise location of the $^6$Li broad Feshbach resonance near 832.2 G, which is constrained by the zero crossing~\cite{JochimPreciseFeshbach}.

\section{Experiment}

Our experiments employ mixtures of the ground Zeeman-hyperfine states of $^6$Li, which are denoted by $|1\rangle$ to $|6\rangle$, in order of increasing energy~\cite{DuSpinSeg1}. We initially prepare a degenerate sample in state $|2\rangle$~\cite{SupportOnline}. The bias magnetic field is tuned to $B=527$ G, near the zero crossing of the $|1\rangle-|2\rangle$ scattering length. A 2 ms radio-frequency $\pi/2$ pulse, which is resonant for transitions from state $|2\rangle$ to state $|1\rangle$ then creates a $|1\rangle-|2\rangle$ superposition state.  Similarly, $|2\rangle-|3\rangle$ or $|1\rangle-|3\rangle$ superposition states are prepared close to the corresponding zero crossings near 589 G or 569 G~\cite{SupportOnline}. The curvature of the bias magnetic field, $B_z(x)$, creates a significant spin-dependent harmonic potential in the long x-direction of the cigar-shaped cloud, with negligible effect in the narrow transverse directions.

The subsequent evolution of the observed spin densities, Fig.~\ref{fig:Clouds}, can be understood using a Bloch vector picture~\cite{DuSpinSeg2}. First, the short radio-frequency (rf) pulse creates a collective spin vector along one axis in the x-y plane.  In a frame rotating about the $z$-axis at the resonant hyperfine frequency,  spin vectors for atoms in the $n^{th}$ axial harmonic oscillator state precess about the $z$-axis at the detuning frequency, $\Omega (E)=-n(E)\,\delta\omega_x$. Here,  $\delta\omega_x=\omega_{x2}-\omega_{x1}=-2\pi\times 14.9\times 10^{-3}$ Hz is the difference in the oscillation frequencies of states $|1\rangle$ and $|2\rangle$, arising from magnetic field curvature, and $n(E)\simeq E/\hbar\bar{\omega}_x$, with $\bar{\omega}_x= (\omega_{x2}+\omega_{x1})/2=2\pi\times 23.0$ Hz. For typical conditions, $E_F=0.56\,\mu$K, the detuning for the average x-energy, $\bar{E}=E_F/4$, is $\Omega(\bar{E})\simeq - 2\pi\times 2.0$ Hz. After coherent excitation, $\Omega(E)$ causes the spin vectors for atoms of different energies to fan out in the $x-y$ plane. Second, forward s-wave scattering, which is not Pauli-blocked in degenerate samples, occurs between two atoms with different energies and corresponding spin vectors, producing a rotation about the total spin vector~\cite{Levitov,Fuchs,Williams,LaloeSpinReph}. This creates a mean field rotation  of the collective spin with  an energy-dependent z-component, which maps into a spatially varying spin density in the harmonic trap, revealed using absorption imaging of both hyperfine components, as shown in Fig.~\ref{fig:Clouds}. The evolution occurs on a time scale set by the mean field frequency, $\Omega_{MF}\simeq 2\pi\times 1.0$ Hz, as discussed below.

\begin{figure*}[htb]
\begin{center}\
\includegraphics[width=6.1in]{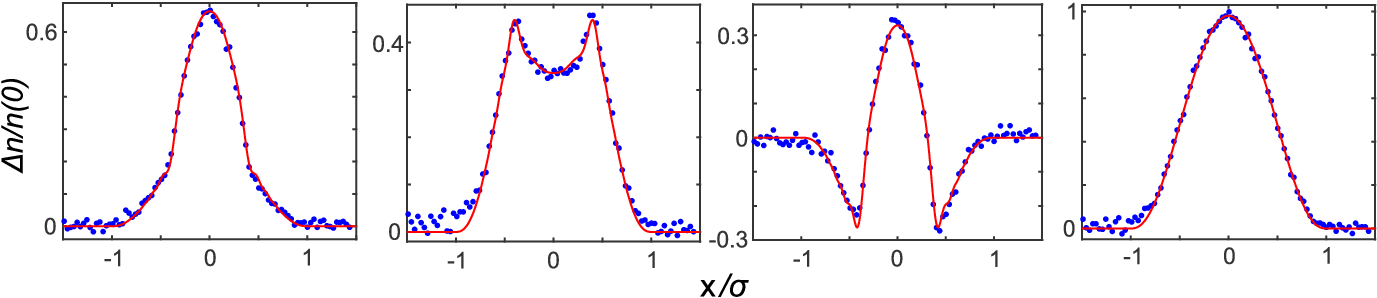}
\end{center}
\caption{Spin-density profiles for a degenerate ($T/T_F=0.35$) Fermi gas at $t=800$ ms relative to coherent excitation. Data (blue dots) versus prediction (red curves) showing quantitative agreement. Left to right: $n_1$, $n_2$, $n_1-n_2$, $n_1+n_2$ in units of the peak total density. Each solid curve is the mean field model with a fixed scattering length of $a=3.04$ bohr ($B=528.147$ G) and a fitted cloud size $\sigma_{Fx}\equiv\sigma=329\,\mu$m, obtained by fitting the total density $n_1+n_2$  to a  1D Thomas-Fermi profile, eq.~\ref{eq:TFdensity}.
\label{fig:4profilesA}}
\end{figure*}

\begin{figure*}[tbh!]
\begin{center}\
\includegraphics[width=6.2in]{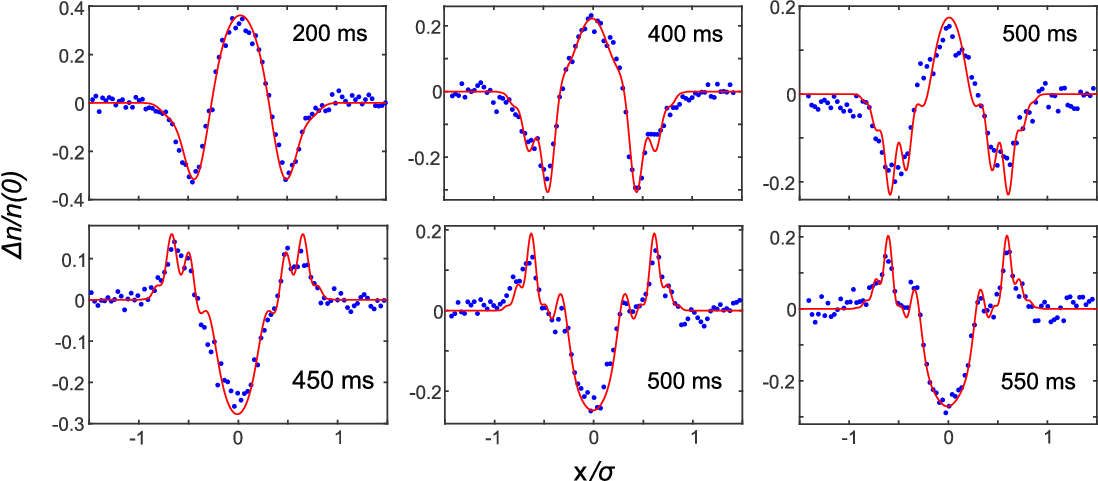}
\end{center}
\caption{Spin-density profiles in a degenerate sample $T/T_F=0.35$ at selected times relative to coherent excitation. $\Delta n(0)=n_1(0,t)-n_2(0,t)$ is given in units of $n_1(0)+n_2(0)$. Solid curves: Mean field model with the same scattering length for each time and a fitted cloud size within a few percent of the measured average value, $\sigma=322.0(1.5)\,\mu$m. Top three panels: $B=528.817$ G, $a=5.17\,a_0$. Bottom three panels: $B=525.478$ G, $a=-5.39\,a_0$. Note that the spin density inverts when the scattering length changes sign.
\label{fig:SpinDensityProfiles}}
\end{figure*}

Fig.~\ref{fig:4profilesA} shows the transversely integrated spin densities 800 ms after coherent excitation, for a degenerate $|1\rangle-|2\rangle$ cloud with $a=3.04\,a_0$.  Fig.~\ref{fig:SpinDensityProfiles} shows the difference of the transversely integrated  spin densities $n_1(x,t)-n_2(x,t)\equiv 2S_z(x,t)$ at selected times $t$ after excitation, for scattering lengths of larger magnitude, $\simeq\pm 5\,a_0$. For the larger scattering lengths, the data are sensitive to the evolution time and exhibit a complex structure, which we explain using a mean field model, outlined below~\cite{SupportOnline}.

\begin{figure*}[htb]
\begin{center}\
\includegraphics[width=6.8in]{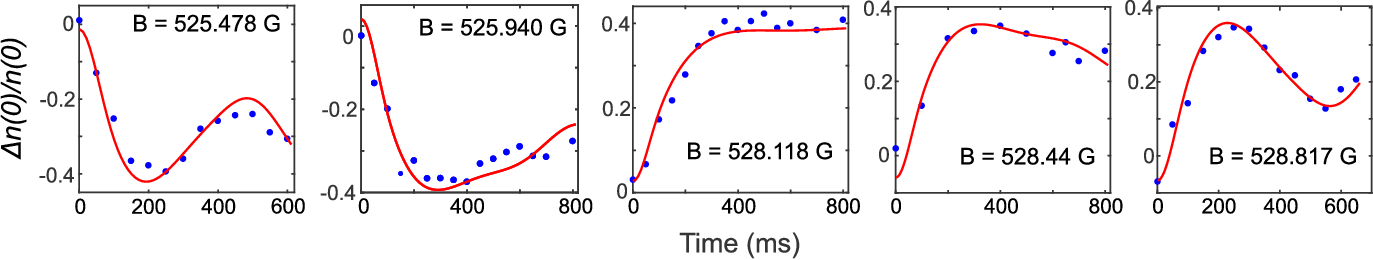}
\end{center}
\caption{Central spin density versus evolution time for various magnetic fields near the zero crossing of the $|1\rangle-|2\rangle$ scattering length. $\Delta n(0)=n_1(0,t)-n_2(0,t)$ is given in units of $n_1(0)+n_2(0)$. Solid curves show the mean-field model with the scattering length $a$ as a fit parameter. The fitted values of $a$ are plotted in Fig.~\ref{fig:AvsB12}.
\label{fig:amplitude}}
\end{figure*}

A thermal average of the Heisenberg equations for the collective spin vector $\tilde{\mathbf{S}}(E,t)$ as a function of axial energy $E$ (in a one dimensional approximation) yields~\cite{DuSpinSeg2,SupportOnline},
\begin{equation}
\partial_t\tilde{\mathbf{S}}(E)={\mathbf{\Omega}}(E)\!\times\tilde{\mathbf{S}}(E)
+\!\!\int\!\!dE'\tilde{g}(E'\!\!,\!E)\,\tilde{\mathbf{S}}(E')\!\times\tilde{\mathbf{S}}(E),
\label{eq:19.6a}
\end{equation}
where we suppress $t$ in $\tilde{\mathbf{S}}(E,t)$ and $\tilde{\mathbf{S}}(E',t)$.
In eq.~\ref{eq:19.6a}, $\mathbf{\Omega}(E,t)$ includes the energy dependent precession rate about the $z$ axis and a general Rabi vector for radio frequency (rf) excitation of the initial superposition state, with $\int dE\, \tilde{S}_z(E,t=0)=1$, prior to the rf pulse. The integral term describes the rotation of the spin vector for atoms of energy $E$ arising from collisions with atoms of energy $E'$. Here, the coupling matrix $\tilde{g}(E',E)$ (see eq.~\ref{eq:g}) is proportional to the mean field frequency and plays the role of the site to site coupling in a lattice model.

For a degenerate gas, the mean field frequency $\Omega_{MF}=9\hbar\,n_{3D}\,a/(5\,m)$, where $n_{3D}$ is the 3D total atom density and $m$ is the atom mass. Although it is not necessary to make a continuum approximation, in eq.~\ref{eq:19.6a} we have assumed that the harmonic oscillator states are closely spaced compared to the Fermi energy, as is the case for our experiments. Employing a WKB approximation for the harmonic oscillator wave functions, $\tilde{g}(E',E)$ is proportional to $1/\sqrt{E-E'}$, which determines the effective long-range character of the spin couplings. Eq.~\ref{eq:19.6a} is solved numerically for $\tilde{\mathbf{S}}(E,t)$, from which we obtain the vector spin density as a function of axial position $x$,
\begin{equation}
{\mathbf{S}}(x,t)=\frac{N}{2}\frac{\bar{\omega}_x}{\pi}\int_0^\infty dp_x\,\tilde{\mathbf{S}}\left(\frac{p_x^2}{2m}+\frac{m\bar{\omega}_x^2}{2}x^2,t\right).
\label{eq:spindensity1A}
\end{equation}

For  $|a|\sim 5$ bohr, with the parameters for our experiments, $\Omega_{MF}\simeq 2\pi\times 1.0$ Hz~\cite{SupportOnline}, while the collision rate~\cite{GehmCollisionRate} is 0.004 $s^{-1}$, which is negligible. As $N_1(E)+N_2(E)$ is conserved \cite{SupportOnline}, the total atom spatial density, determined by analogy to eq.~\ref{eq:spindensity1A}, should be constant in time, as shown in Fig.~\ref{fig:Clouds}.

For the low temperature, degenerate gas,  we find that eq.~\ref{eq:19.6a} is in excellent quantitative agreement with the spin-density profiles of Fig.~\ref{fig:4profilesA} and captures very well the fine features of the data shown in Fig.~\ref{fig:SpinDensityProfiles}, as well as the time dependence of the spin-density profiles shown in  Fig.~\ref{fig:TimeDepSpinDensity} for a fixed scattering length~\cite{SupportOnline}.

We fit the mean field model to the data of Fig.~\ref{fig:SpinDensityProfiles} in the following way. First, we plot the dimensionless spin density  $(n_1-n_2)/(n_1+n_2)$ at the center ($x=0$) as a function of time,  Fig.~\ref{fig:amplitude}, for each value of the magnetic field. Second, we fit the model to the data of Fig.~\ref{fig:amplitude}  to find  the scattering length that gives the best fits (red curves). The fits to the spatial density profiles of are then obtained by fixing the scattering length at each field to the value obtained from Fig.~\ref{fig:amplitude} and adjusting the Thomas-Fermi radius by a few per cent to fit the measured profile at each time. The mean of the measured radii is found to be 322.0(1.5) $\mu$m. Magnetic field stability is better than $5$ mG, limited by measurement precision. The absolute value of the field is calibrated using radio frequency spectroscopy of the hyperfine transitions.

\begin{figure*}[htb]
\begin{center}\
\includegraphics[width=3.8in]{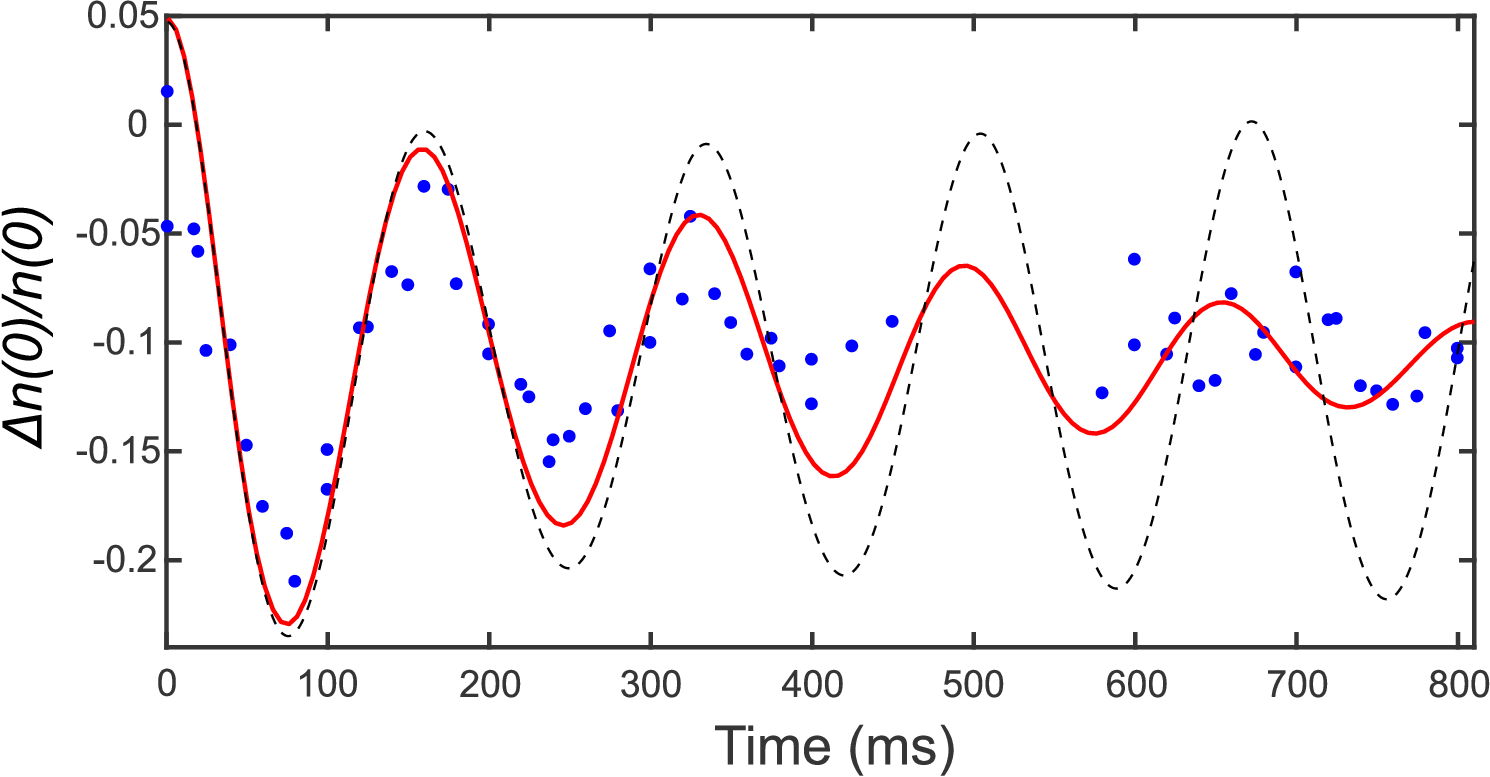}
\end{center}
\caption{Decay of the amplitude of the central spin density versus time for $a=-14.9\,a_0$. The dashed curve shows the predicted amplitude for the average density. The red curve shows the the average of the predictions based on the measured atom numbers and cloud widths for each shot.
\label{fig:oscampldecay}}
\end{figure*}

\begin{figure*}[htb]
\begin{center}\
\includegraphics[width=5.3in]{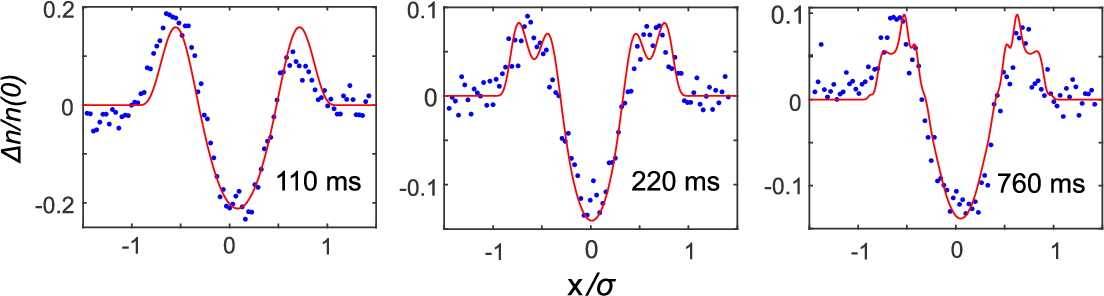}
\end{center}
\caption{Spin density profiles  versus time for $a=-14.9\,a_0$ versus predictions (red curves) with the same scattering length for each time and a fitted cloud size within a few percent of the measured average value, $\sigma=330.6 \,\mu$m.
\label{fig:decayspinprofiles}}
\end{figure*}

Increasing the scattering length to $a=-14.9\,a_0$, we measure the amplitude of the spin density at the cloud center  for a degenerate sample as a function of time relative to coherent excitation, Fig.~\ref{fig:oscampldecay}. Although the collision rate $\simeq 0.04\,s^{-1}$ is still negligible,
we observe a decay of the amplitude that is not predicted.  We believe that the decay arises from the variation of the atom density over several runs, which are averaged to determine each data point. The average of the predictions (red curve) of Fig.~\ref{fig:oscampldecay} yields the observed decay, because the sensitivity to the mean field frequency, and hence to the atom density variation, increases with increasing time, resulting in a decreasing amplitude for the average. The corresponding spatial profiles are shown in Fig.~\ref{fig:decayspinprofiles}, where predicted curves are obtained for a fixed scattering length of $-14.9\,a_0$ and fitting the Fermi width, within a few percent of the mean.

\section{Scattering Length Parameters}

The small $a$ region, where the mean field model precisely fits the data, enables measurement of the tuning rate $a'$ (in bohr per gauss) of the scattering length near the zero crossing field $B_0$, where
\begin{equation}
a(B)=a'\,(B-B_0).
\label{eq:scattlength1}
\end{equation}
Here, we assume that the energy shift is negligible for the degenerate sample, in contrast to the hot sample discussed  below. Using the data in Fig.~\ref{fig:amplitude}, the fitted $|1\rangle-|2\rangle$ scattering length  for each magnetic field is plotted in Fig.~\ref{fig:AvsB12}. The corresponding plot for $|2\rangle-|3\rangle$ scattering is discussed in Appendix~\ref{sec:ExptMethods}. The slopes of the linear fits to the data yield the tuning rates $a'$, Table I.

\begin{figure}[htb]
\begin{center}\
\includegraphics[width=2.5in]{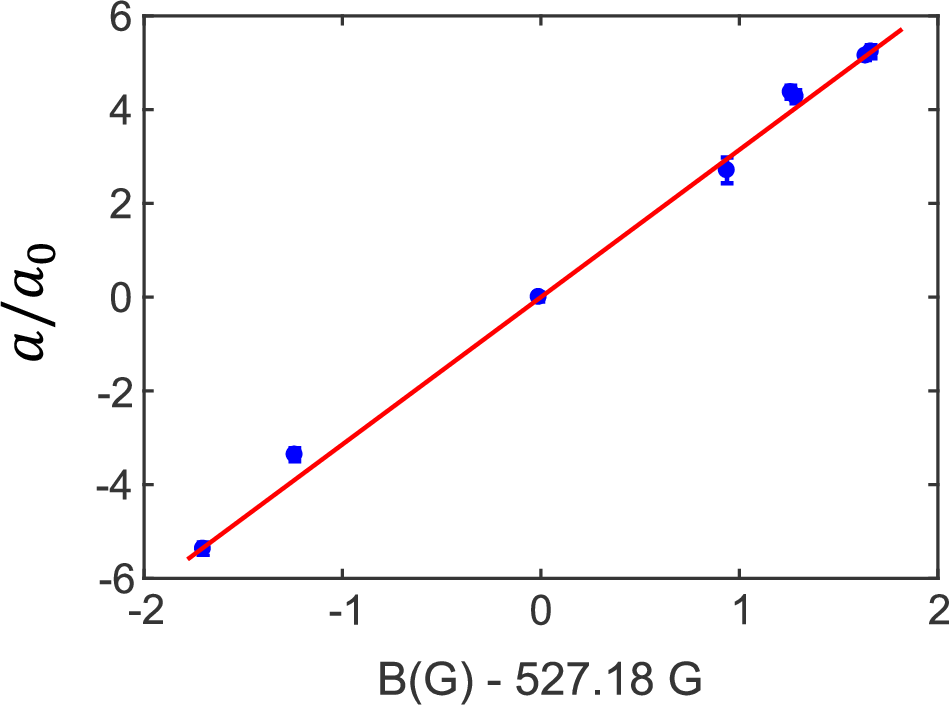}
\end{center}
\caption{Fitted scattering length $a$ versus measured magnetic field for a $|1\rangle-|2\rangle$ mixture ($a_0=1$ bohr). Error bars denote one standard deviation, obtained for each $\chi^2$ fit of Fig.~\ref{fig:amplitude}.
\label{fig:AvsB12}}
\end{figure}

\begin{figure}[htb]
\begin{center}\
\includegraphics[width=2.85in]{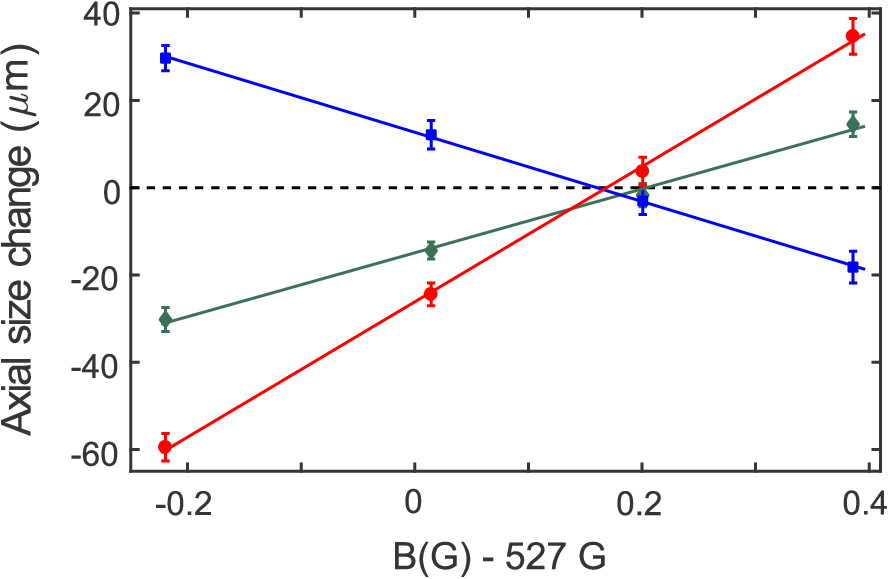}
\end{center}
\caption{Measurement of the zero crossing field for a degenerate $^6$Li $|1\rangle-|2\rangle$ mixture. The plots show the change in cloud size between $t=0$ and $t=800$ ms for state 1 (squares), state 2 (diamonds), and the difference in the cloud sizes of the two spin states at $t=800$ ms (circles). Solid lines are corresponding linear fits, crossing zero (dashed line) when $a=0$. Error bars denote the standard deviation of the mean of five runs.
\label{fig:crossing}}
\end{figure}

Next, we measure the magnetic field $B_0$ at which the scattering length vanishes by using the spin evolution as a sensitive probe: The profiles of the individual spin components remain unchanged at the zero crossing in the degenerate regime. Fig.~\ref{fig:crossing} shows the change in size for each spin profile between $t=0$ and $t=800$ ms, as a function of magnetic field. In addition, we show the difference between the sizes of the state 1 and state 2 profiles at $t=800$ ms. Each method gives a field value $B_0$ for the zero crossing. We report the mean in Table I. The corresponding uncertainties are estimated as one half of the difference between the maximum and the minimum of $B_0$.
\begin{table*}
\caption{Zero crossings $B_0$(G) and tuning rates $a'(a_0/{\rm G})$ for the scattering lengths of the broad Feshbach resonances in $^6$Li.  \label{Feshbach1}}
\vspace*{0.25in}
\begin{tabular}{|c|c|c|c|c|c|c|c|}
            \hline
             States&T($\mu$K)&\hspace{0.1in}$B_0$(G)~[This work]\hspace{0.25in}&$B_0$(G)~\cite{BartensteinFeshbach}&$B_0$(G)~\cite{JochimPreciseFeshbach}
             &\hspace*{0.05in}$a'(a_0/G)$~[This work]&$a'(a_0/G)$\cite{BartensteinFeshbach}&$a'(a_0/G)$~\cite{JochimPreciseFeshbach}\\
             \hline
             1-2  &0.2& 527.18(2)  &543.15& 527.32(25)&3.14(8) &4.12 &3.49\\
             1-2  &45.7& 527.42(1) &- &- &- &- &-\\
             2-3    &0.2&588.68(1) & 588.92& 588.75&4.52 (23) &6.11 &5.82\\
			 1-3    &0.2&567.98(1) & 568.13& 568.02&- &13.87 &13.29 \\
             \hline
\end{tabular}
\end{table*}
The zero crossing for $a_{12}$, 527.18(2) G, is smaller than the value 527.5(2) G obtained by the same method at high temperature~\cite{DuSpinSeg1}, and is consistent with the calculated value 527.32(25), based on the most recent $^6$Li$_2$ molecular potentials determined from 1D dimer spectra~\cite{JochimPreciseFeshbach}. The zero crossings for $a_{13}$, 567.98(01) G and for $a_{23}$, 588.68(01), listed Table~\ref{Feshbach1}, are in very good agreement with the values 568.07 G and 588.80 G estimated from the Feshbach resonance data of Ref.~\cite{JochimPreciseFeshbach}, which differ only slightly from Ref.~\cite{BartensteinFeshbach}.

Table~\ref{Feshbach1} compares that the tuning rates  $a_{12}'=3.14\,a_0/$G and $a_{23}'=4.52\,a_0/$G, which we obtain from the fitted scattering length versus magnetic field in the present work, to estimates based on the Feshbach resonance profiles $a[B]$, which are obtained from the molecular potentials reported in Ref.~\cite{BartensteinFeshbach} and in Ref.~\cite{JochimPreciseFeshbach,Julienne}. Using the profiles of Ref.~\cite{BartensteinFeshbach}, we find $a_{12}'=4.12\,a_0/$G and $a_{23}'=6.11\,a_0/$G. These slopes are 50\% larger than those estimated in the present work, but the ratios, $4.52/3.14=1.44$ and $6.11/4.12=1.48$, are in good agreement. This suggests that the discrepancy may be explained by an overall scale factor in our estimate of the transverse averaged 3D density $n_{3D}$ (see eq.~\ref{eq:nperpav}), which determines the scattering lengths from the mean field frequencies $\Omega_{MF}\propto n_{3D}\,a$ used to fit Fig.~\ref{fig:amplitude}. However, using the Feshbach resonance profiles of Ref.~\cite{JochimPreciseFeshbach,Julienne}, we estimate the tuning rate $a_{12}'=3.51\,a_0/$G, which only 11\% larger than tuning rate obtained from our experiments, and $a_{23}'=5.82\,a_0/$G, which is  29\% larger.

\section{Energy Shift}

We also observe the energy dependent shift in the zero crossing, by preparing a $|1\rangle-|2\rangle$ superposition at a high temperature of $T=45.7\,\mu$K. There, we measure a shift of $0.22$ G relative to the degenerate sample. This yields an energy tuning rate of $4.7$ mG/$\mu$K, confirming that the energy dependent shift is negligible for the degenerate samples, compared to the precision of the magnetic field measurement.

\begin{figure}[htb]
\begin{center}\
\includegraphics[width=2.23in]{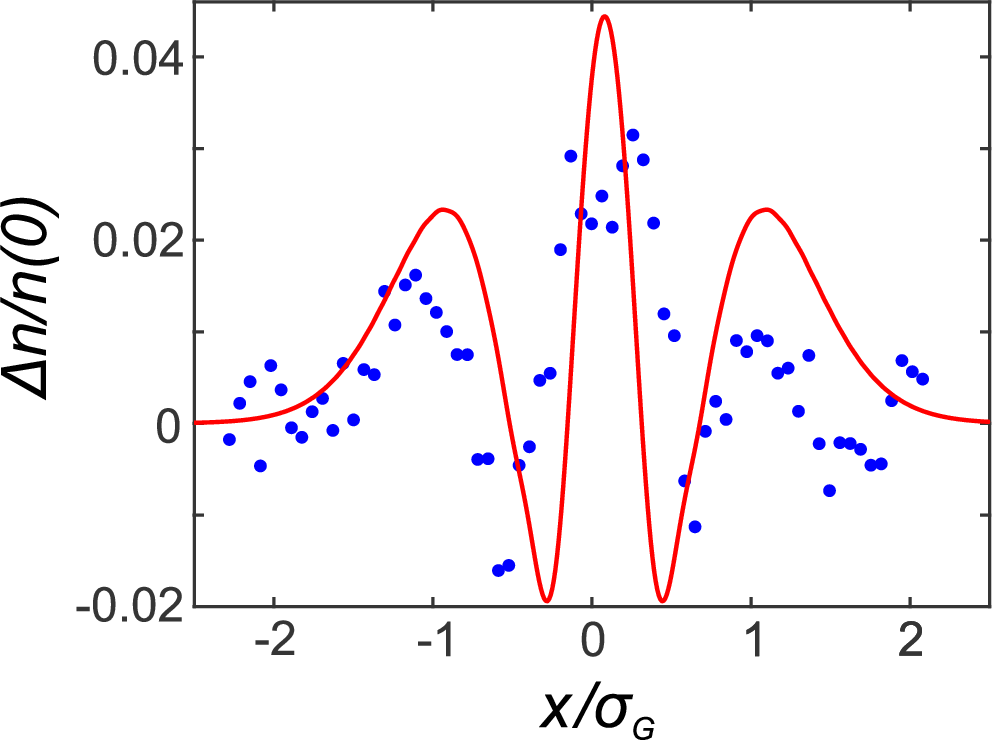}
\end{center}
\caption{High temperature spin density profile of a $|1\rangle-|2\rangle$ mixture for $t=400$ ms. $T=45.7\,\mu$K and $B=527.466$ G, where the zero-energy s-wave scattering length is $0.90$ bohr. Here $\sigma_G=323\,\mu$m is the gaussian $1/e$ radius of the total density profile.
\label{fig:HighTSmalla}}
\end{figure}

To directly illustrate the energy dependence, we  measure the spin density at $45.7\,\mu$K for $B=527.466$ G, Fig.~\ref{fig:HighTSmalla}.
We see that the high temperature spin density profile crosses the zero axis four times, in contrast to the low temperature data of Fig.~\ref{fig:SpinDensityProfiles}, which only crosses twice.

The modification of the spin-density profile at high temperature is not likely to arise from the $|1\rangle-|2\rangle$ p-wave resonance in $^6$Li, which is located near 186.2(6) G and has a width of $0.5$G~\cite{SalomonPWave}. To understand this profile and the energy shift, we include the energy dependence of the scattering length and of the average magnetic field, by replacing  $a$ in $\tilde{g}(E',E)$ of eq.~\ref{eq:19.6} with $a(E',E)=a'[B_{\rm eff}(E',E)-B_0]$~\cite{SupportOnline}, with $B_{\rm eff}(E',E)$ the effective magnetic field. Then, for small positive $B-B_0$, atoms with small energies $E,E'$ have $B_{\rm eff}-B_0>0$ and a positive  scattering length, while  atoms with high energies $E,E'$, have $B_{\rm eff}-B_0<0$ and a negative scattering length. These two contributions result in the extra crossings. The solid red curve of Fig.~\ref{fig:HighTSmalla} includes  the average transverse kinetic energy, which shifts the effective field from the applied value of $0.28$ G above the zero crossing to $0.08$ G above, where $a=0.25\,a_0$ for atoms with $E=E'=0$.

In summary, we have shown that a mean field collective spin rotation model, including the full energy-dependent coupling matrix, quantitatively describes the spin density evolution in the collisionless regime, precisely testing the underlying energy-space spin-lattice model. The measurements provide an essential benchmark for future work on collective spin evolution with designer energy landscapes in the weakly interacting regime, and pave the way studies of beyond mean field physics in weakly interacting gases, measurement of spatially correlated spin fluctuations~\cite{KollerReySpinDep} and measurement of correlated spin currents~\cite{DuineMagnonSpinCurrent}.

Primary support for this research is provided by the Physics Divisions of the Army Research Office (W911NF-14-1-0628) and the Division of Materials Science and Engineering, the Office of Basic Energy Sciences, Office of Science, U.S. Department of Energy (DE-SC0008646). Additional support for the JETlab atom cooling group has been provided by the National Science Foundation (PHY-1705364) and the Air Force Office of Scientific Research (FA9550-16-1-0378).\\

$^*$Corresponding author: jethoma7@ncsu.edu
\newpage
%

\appendix

\section{Experimental Methods}
\label{sec:ExptMethods}

A cloud comprising a 50-50 mixture of the two lowest hyperfine states, denoted  $|1\rangle$ and $|2\rangle$, is evaporatively cooled to degeneracy near the $|1\rangle-|2\rangle$ Feshbach resonance at $832.2$ G. The magnetic field is then ramped to the weakly interacting regime near 1200 G, and the $|1\rangle$ spin component is eliminated by means of a resonant optical pulse. To create a $|1\rangle-|2\rangle$ superposition state, the magnetic field is ramped to 527 G, near the zero crossing of the scattering length. The atoms in spin state $|2\rangle$ are then excited by a 2 ms radio-frequency $\pi/2$ pulse, which is resonant for transitions to state $|1\rangle$. Similarly, a $|2\rangle-|3\rangle$ superposition state is prepared by employing an rf transition from state $|2\rangle$ to state $|3\rangle$ close to the corresponding zero crossing around 589 G. For the $|1\rangle-|3\rangle$ superposition state, we prepare a single $|2\rangle$ spin component at 1200 G. The magnetic field is then ramped down to the value of interest around 568 G, near the zero crossing of the $|1\rangle-|3\rangle$ scattering length. The atoms are excited by a 2 ms radio-frequency $\pi/2$ pulse, which is resonant with the transition from state $|2\rangle$ to state $|1\rangle$, creating a balanced $|1\rangle-|2\rangle$ superposition state. Then a 4 ms radio-frequency $\pi$ pulse is applied, which is resonant with the transition from state $|2\rangle$ to state $|3\rangle$, to create a balanced $|1\rangle-|3\rangle$ superposition state. The trap parameters for our experiments are: $\omega_{\rm mag}^2=(2\pi\times 20.5\,{\rm Hz})^2\,{\rm B(G)}/834$; $\bar{\omega}_x=2\pi\times 23$ Hz, $\omega_\perp=2\pi\times 625$ Hz, for the degenerate gas, and $\bar{\omega}_x=2\pi\times 174$ Hz, $\omega_\perp=2\pi\times 5.77$ kHz, for the high temperature gas.

After preparation, we obtain degenerate samples with a typical total atom number  of  $N=N_\uparrow+N_\downarrow\simeq 7.0\times 10^4$ and an ideal gas Fermi temperature of $k_BT_F=\hbar(3N\bar{\omega}_x\omega_\perp^2)^{1/3}=k_B\times 0.56\,\mu$K for our trap frequencies.  To determine the temperature $T$, the measured one dimensional total density versus $x$ is fit with a finite temperature Thomas-Fermi profile for a noninteracting gas, which is appropriate for our weakly interacting gas.  Using the calculated Thomas-Fermi radius $\sigma_{TF}=\sqrt{2\,k_BT_F/(m\bar{\omega}_x^2)}= 270\,\mu$m, we find $T=0.35\,T_F$.

\begin{figure}[htb]
\begin{center}\
\includegraphics[width=2.5in]{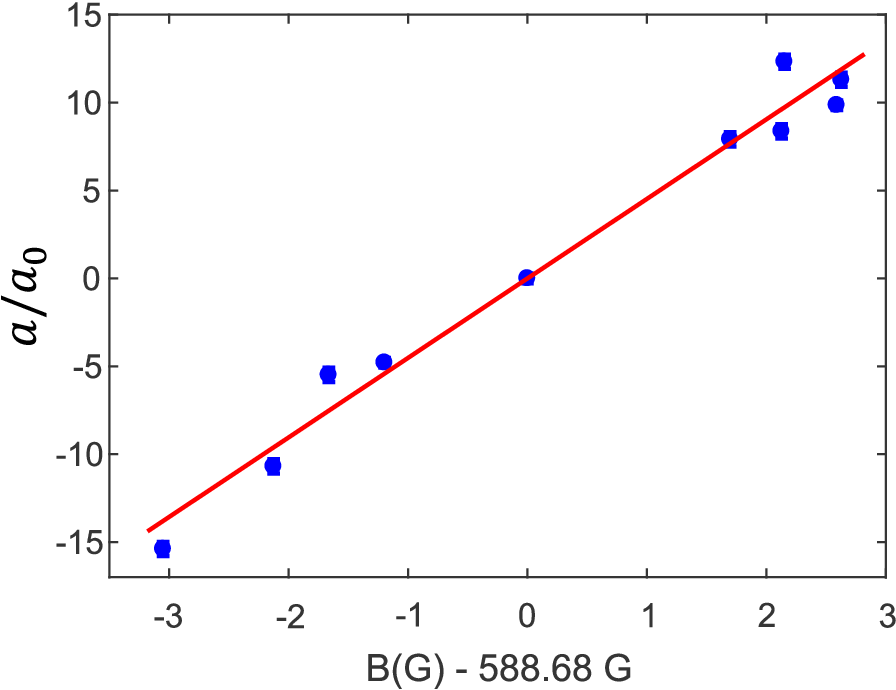}
\end{center}
\caption{Tuning rate of the scattering length $a$ of a $|2\rangle-|3\rangle$ mixture versus measured magnetic field ($a_0=1$ bohr) Error bars denote one standard deviation, obtained for each $\chi^2$ fit to the time dependent central amplitude for the given $B$.
\label{fig:AvsB23}}
\end{figure}

In the main text, we reported measurements of the zero crossing field of the scattering length for $^6$Li $|1\rangle-|2\rangle$, $|2\rangle-|3\rangle$, and $|1\rangle-|3\rangle$ mixtures and the tuning rate of the scattering length for $|1\rangle-|2\rangle$ and $|2\rangle-|3\rangle$ mixtures. Fig.~\ref{fig:AvsB23} shows the data that was used to obtain the tuning rate for the $|2\rangle-|3\rangle$ mixture.

\begin{figure}[htb]
\begin{center}\
\includegraphics[width=2.5in]{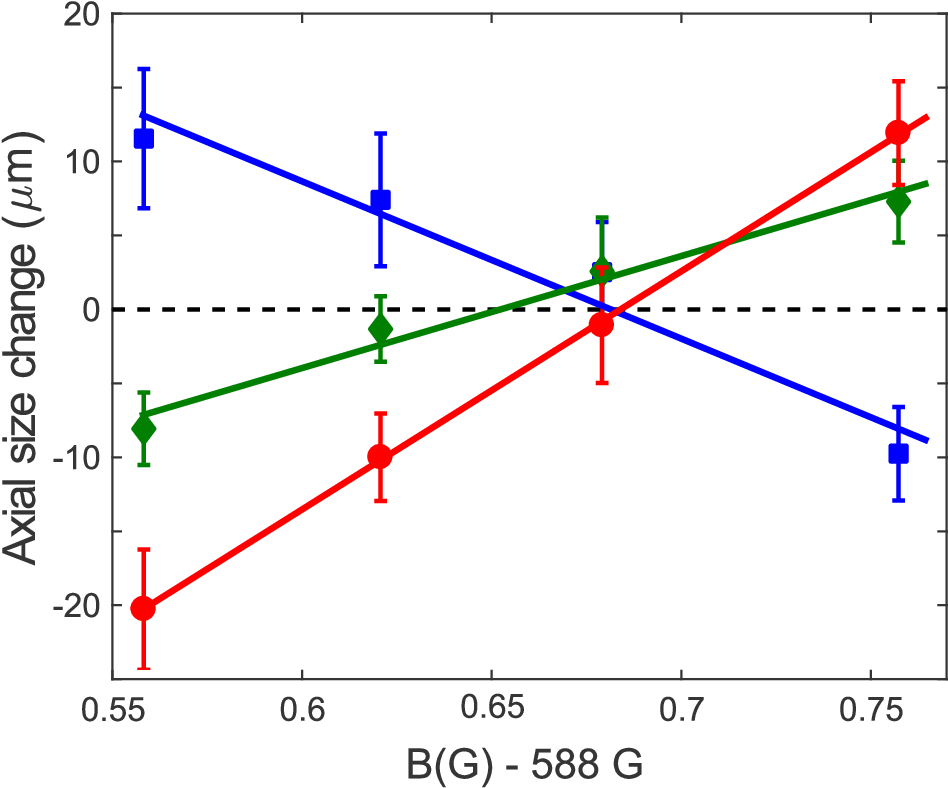}
\end{center}
\caption{Measurement of the zero crossing field for a degenerate $^6$Li $|2\rangle-|3\rangle$ mixture. The plots show the change in cloud size between $t=0$ and $t=800$ ms for state 3 (squares), state 2 (diamonds), and the difference in the cloud sizes of the two spin states at $t=800$ ms (circles). Solid lines are corresponding linear fits, crossing zero (dashed line) when $a=0$. Error bars denote the standard deviation of the mean of five runs.
\label{fig:crossing23}}
\end{figure}

\begin{figure}[htb]
\begin{center}\
\includegraphics[width=2.5in]{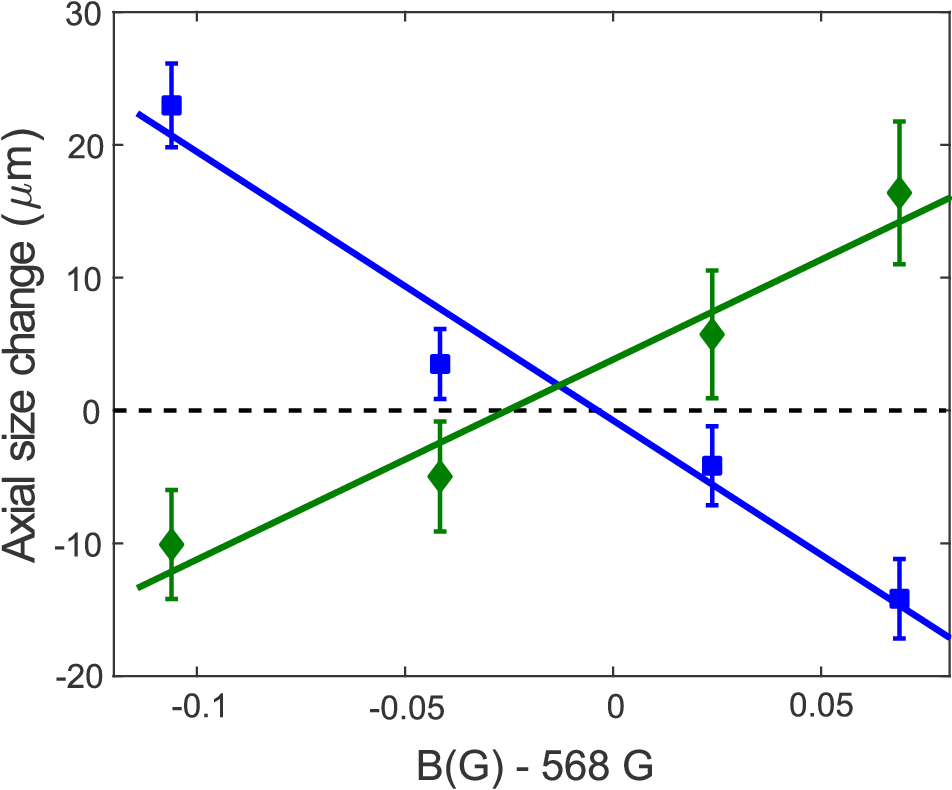}
\end{center}
\caption{Measurement of the zero crossing field for a degenerate $^6$Li $|1\rangle-|3\rangle$ mixture. The plots show the change in cloud size between $t=0$ and $t=800$ ms for state 3 (squares), state 1 (diamonds). Solid lines are corresponding linear fits, crossing zero (dashed line) when $a=0$. Error bars denote the standard deviation of the mean of five runs.
\label{fig:crossing13}}
\end{figure}

Figs.~\ref{fig:crossing23}~and~\ref{fig:crossing13} show the data that was used to obtain the zero crossings fields for the $|2\rangle-|3\rangle$, and $|1\rangle-|3\rangle$ mixtures. In Fig.~\ref{fig:crossing23} and Fig.~\ref{fig:crossing13}, and for Fig.~5 of the main paper, we take into account cloud size variations arising from small changes in the atom number. Each data point represents an average of $10$ experimental runs. For each run $i$, we extract the atom number $N_i$ and the axial cloud size $\sigma_i$ for each spin component. The cloud sizes scale as $N_i^{1/6}$ for zero temperature Thomas-Fermi profiles. Therefore, to correct for the varying atom number, we calculate the reduced size $\sigma_i/N_i^{1/6}$  for each run and use $\left\langle \sigma_i/N_i^{1/6} \right\rangle \left\langle N_i^{1/6}\right\rangle$ as the effective mean cloud size for each field.

\section{Mean-Field Model}
\label{sec:MFmodel}

 We employ a mean field model in energy representation to describe the spin-density profiles observed in our experiments. The bias magnetic field tunes the s-wave scattering length near the zero crossing, where the gas is very weakly interacting and the energy changing collision rate is negligible. For this reason, we begin with the single particle Hamiltonian for a noninteracting Fermi gas with two spin states, a lower hyperfine state denoted $\uparrow$ and an upper hyperfine state denoted $\downarrow$. For an atom at rest, these states differ in energy by $\hbar\omega_{HF}$, where $\omega_{HF}$ is the hyperfine resonance frequency. A spin-independent cigar-shaped optical trap confines the atom cloud weakly along the cigar axis, denoted $x$, and tightly in the perpendicular $\rho$ direction, so that $\rho<<|x|$.  Curvature in the bias magnetic field produces a significant harmonic confining potential along the $x$-axis, while for the $\rho$ direction, the magnetic contribution to the confining  potential is negligible compared to that of the optical trap. The net optical and magnetic trapping potential along $x$ is then spin-dependent, with harmonic oscillation frequencies $\omega_{x\uparrow}$ and $\omega_{x\downarrow}$. The Hamiltonian for the motion along the $x$-axis (without the hyperfine energies) is
\begin{eqnarray}
H_0&=&\sum_n \ket{n}\bra{n}\Big[(n+1/2)\,\hbar\omega_{x\uparrow}\ket{\uparrow}\bra{\uparrow}\nonumber\\
& &\hspace{0.25in}+(n+1/2)\,\hbar\omega_{x\downarrow}\ket{\downarrow}\bra{\downarrow}\Big].
\label{eq:1.1}
\end{eqnarray}
For later use, we define the dimensionless single particle spin operators,
\begin{eqnarray}
s_z&=&\frac{\ket{\uparrow}\bra{\uparrow}-\ket{\downarrow}\bra{\downarrow}}{2}\nonumber\\
s_x&=&\frac{\ket{\uparrow}\bra{\downarrow}+\ket{\downarrow}\bra{\uparrow}}{2}\nonumber\\
s_y&=&\frac{\ket{\uparrow}\bra{\downarrow}-\ket{\downarrow}\bra{\uparrow}}{2i},
\label{eq:2.7}
\end{eqnarray}
where $[s_x,s_y]=s_xs_y-s_ys_z=i s_z$ and cylic permutations.

A radio-frequency transition does not change the harmonic oscillator quantum number $n$. Hence, the resonance frequency for a transition from the lower $\uparrow$ to the upper $\downarrow$ hyperfine state of an oscillating atom in state $\ket{n}$ is $\omega_{res}=\omega_{HF}+(n+\frac{1}{2})\,\delta\omega_x$ with $\delta\omega_x\equiv \omega_{x\downarrow}-\omega_{x\uparrow}$. Working in a frame rotating at the hyperfine resonance frequency $\omega_{HF}$ and
defining the energy $E=(n+\frac{1}{2})\,\hbar\bar{\omega}_{x}$, where the mean oscillation frequency, $\bar{\omega}_x\equiv (\omega_{x\uparrow}+\omega_{x\downarrow})/2$, we can  rewrite Eq.~\ref{eq:1.1} as
\begin{eqnarray}
H_0&=&\sum_E\ket{E}\bra{E}\Bigg[ E(\ket{\uparrow}\bra{\uparrow}+\ket{\downarrow}\bra{\downarrow})\nonumber\\
& &\hspace{0.25in}+\,\hbar\Omega(E)\,\frac{\ket{\uparrow}\bra{\uparrow}-\ket{\downarrow}\bra{\downarrow}}{2}\Bigg],
\label{eq:18}
\end{eqnarray}
where $\langle E'\ket{E}=\delta_{E',E}$ and  the last term is proportional to $s_z$, with
\begin{equation}
\Omega(E)\equiv -\delta\omega_x\,\frac{E}{\hbar\bar{\omega}_x}.
\label{eq:1.7}
\end{equation}

To treat the many-body problem for a very weakly interacting gas, where the single particle energies do not change during the evolution time, we define the field operator in energy representation,
\begin{equation}
\hat{\psi}\equiv \sum_{E,\sigma=\uparrow,\downarrow}\hat{a}_\sigma(E)\,\ket{E}\ket{\sigma}.
\label{eq:2.1}
\end{equation}
 With the anticommutation relations
 \begin{equation}
 \{\hat{a}_\sigma(E),\hat{a}^\dagger_{\sigma'}(E')\} =\delta_{\sigma,\sigma'}\delta_{E,E'},
 \label{eq:2.2}
 \end{equation}
 we have $\{\hat{\psi},\hat{\psi}^\dagger\}=\hat{1}$, the product of the energy and spin identity operators. The many-body Hamiltonian for the noninteracting atoms is then defined by $\hat{H}_0=(\hat{\psi}^\dagger H_0\hat{\psi})$, where the parenthesis $(...)$ denotes inner products for the {\it single particle} energy and spin states, $|E\rangle\,|\uparrow,\downarrow\rangle$. Then,
 \begin{eqnarray}
 \hat{H}_0&=&\sum_{E'}E'[\hat{N}_\uparrow(E')+\hat{N}_\downarrow(E')]\nonumber\\
 & &+\sum_{E'}\hbar\Omega(E')\,\hat{S}_z(E').
 \label{eq:3.3}
 \end{eqnarray}
 Here, the number operators are $\hat{N}_\uparrow(E)=a^\dagger_\uparrow(E)a_\uparrow(E)$ and $\hat{N}_\downarrow(E)=a^\dagger_\downarrow(E)a_\downarrow(E)$ and the dimensionless many-body spin operators are given (in the Schr\"{o}dinger picture) by
 \begin{eqnarray}
 \hat{S}_z(E)&=&(\hat{\psi}^\dagger|E\rangle s_z\langle E|\hat{\psi})=\frac{\hat{N}_\uparrow(E)-\hat{N}_\downarrow(E)}{2} \label{eq:3.4}\\
 \hat{S}_x(E)&=&(\hat{\psi}^\dagger |E\rangle s_x\langle E| \hat{\psi})=\frac{\hat{a}^\dagger_\uparrow(E)\,\hat{a}_\downarrow(E)+\hat{a}^\dagger_\downarrow(E)\,\hat{a}_\uparrow(E)}{2}\nonumber\\
 \hat{S}_y(E)&=&(\hat{\psi}^\dagger |E\rangle s_y\langle E| \hat{\psi})=\frac{\hat{a}^\dagger_\uparrow(E)\,\hat{a}_\downarrow(E)-\hat{a}^\dagger_\downarrow(E)\,\hat{a}_\uparrow(E)}{2\,i}.\nonumber
 \end{eqnarray}

The corresponding field operators in position representation are
\begin{eqnarray}
\hat{\psi}(x)&=&(\bra{x}\hat{\psi})=\sum_{E,\sigma}\hat{a}_\sigma(E)\phi_E(x)\ket{\sigma}\nonumber\\
&\equiv&\sum_\sigma\hat{\psi}_\sigma(x)\ket{\sigma}.
\label{eq:3.5}
\end{eqnarray}
The Schr\"{o}dinger picture operator of the $z$ component of the spin density is then
\begin{eqnarray}
\hat{S}_z(x)&=&(\hat{\psi}^\dagger(x)s_z\hat{\psi}(x))\label{eq:3.6}\\
& &\hspace{-0.5in}=\frac{1}{2}\sum_{E,E'}\phi^*_{E'}(x)\phi_E(x)\left[\hat{a}^\dagger_\uparrow(E')\,\hat{a}_\uparrow(E)
-\hat{a}^\dagger_\downarrow(E')\,\hat{a}_\downarrow(E)\right].\nonumber
\end{eqnarray}
Note that the orthonormality of the $\phi_E(x)$ yields $\int dx\hat{S}_z(x)=\sum_E\hat{S}_z(E)=\hat{S}_z$, the total $z$-component of the spin operator.

For our mean-field treatment, we assume initially that there is no coherence between $E'\neq E$ for a thermal average, i.e., $\langle \hat{a}^\dagger_\uparrow(E')\hat{a}_\uparrow(E)\rangle=\langle\hat{N}_\uparrow(E)\rangle\,\delta_{E',E}$. Then the $z$-component of the c-number spin density is given by
\begin{equation}
S_z(x)\equiv\langle\hat{S}_z(x)\rangle=\sum_E|\phi_E(x)|^2\langle\hat{S}_z(E)\rangle.
\label{eq:3.7}
\end{equation}
Hence, we need only to determine $S_z(E,t)$ to predict the measured $S_z(x,t)$.

Using the anticommutation relations of Eq.~\ref{eq:2.2},  it is easy to evaluate the elementary commutators,
\begin{eqnarray}
&\hspace*{-0.45in}\left[\hat{a}^\dagger_{\sigma_1'}(E_1')\,\hat{a}_{\sigma_1}(E_1),\hat{a}^\dagger_{\sigma'}(E)\right]=
\hat{a}^\dagger_{\sigma_1'}(E_1')\,\delta_{E1,E}\,\delta_{\sigma_1,\sigma'}\label{eq:4.1}\\
&\hspace*{-0.30in}\left[\hat{a}^\dagger_{\sigma_1'}(E_1')\,\hat{a}_{\sigma_1}(E_1),\hat{a}_{\sigma}(E)\right]=
-\hat{a}_{\sigma_1}(E_1)\,\delta_{E1',E}\,\delta_{\sigma_1',\sigma},\nonumber
\end{eqnarray}
which are formally identical to the results obtained for bosons. With eq.~\ref{eq:4.1}, it is straightforward to show that the spin operators of eq.~\ref{eq:3.4} satisfy the usual cyclic commutation relations,
\begin{equation}
[\hat{S}_i(E'),\hat{S}_j(E)]=i\,\epsilon_{ijk}\,\hat{S}_k(E)\,\delta_{E',E}.
\label{eq:Scomm}
\end{equation}
With eq.~\ref{eq:3.3}, the Heisenberg operator equations for the collisionless spin evolution are then
\begin{equation}
\frac{\partial\hat{\mathbf{S}}(E,t)}{\partial t}=\frac{i}{\hbar}\left[\hat{H}_0,\hat{\mathbf{S}}(E,t)\right]={\mathbf{\Omega}}(E,t)\times\hat{\mathbf{S}}(E,t),
\label{eq:5.5}
\end{equation}
where
\begin{equation}
{\mathbf{\Omega}}(E)=\hat{e}_z \Omega(E)
\label{eq:5.4}
\end{equation}
and $\Omega(E)$ is given by eq.~\ref{eq:1.7}. For sample preparation using  radio frequency excitation, eq.~\ref{eq:5.4} is readily generalized to include a time dependent Rabi frequency rotation rate $\Omega_R(t)\,\hat{e}_y$ and an additional time dependent detuning term $\Delta(t)\,\hat{e}_z$, with $\Delta =\omega(t)-\omega_{HF}$ in the rotating frame.

Next, we consider collisional interactions, assuming s-wave scattering between atoms of different spin, which is dominant at low temperature. Short range scattering is modeled by a contact interaction between spin-up and spin-down atoms with an s-wave scattering length $a_S$,
\begin{equation}
H'({\mathbf{x}}_1-{\mathbf{x}}_2)=\frac{4\pi\hbar^2 a_S}{m}\,\delta({\mathbf{x}}_1-{\mathbf{x}}_2)\equiv g\,\delta({\mathbf{x}}_1-{\mathbf{x}}_2),
\label{eq:contactint}
\end{equation}
For the many-body system,
\begin{eqnarray}
\hat{H}'\!\!&=&\!\!\int\!\! \frac{d^3{{\mathbf{x}}_1} d^3{{\mathbf{x}}_2}}{2}\left(\hat{\psi}^\dagger({\mathbf{x}}_2)\hat{\psi}^\dagger({\mathbf{x}}_1)
H'({\mathbf{x}}_1\!-\!{\mathbf{x}}_2)\hat{\psi}({\mathbf{x}}_1)\hat{\psi}({\mathbf{x}}_2)\right)\nonumber\\
& &=g\,\int d^3{\mathbf{x}}\,\hat{\psi}_\uparrow^\dagger({\mathbf{x}})\hat{\psi}_\downarrow^\dagger({\mathbf{x}}) \hat{\psi}_\downarrow({\mathbf{x}})\hat{\psi}_\uparrow({\mathbf{x}}),
\label{eq:6.1}
\end{eqnarray}
where the factor $1/2$ avoids double counting and $\hat{\psi}^2_{\uparrow,\downarrow}({\mathbf{x}})=0$. For simplicity, we initially neglect the dependence of $a_S$ on the relative kinetic energy of the colliding pair, which will be included later.

For our experiments, where atoms are confined in a cigar-shaped cloud, the $x$ dimension is large compared to the radial dimension $\rho$, so that the bias field curvature is negligible along the $\rho$ direction, as noted above.  Therefore, we treat the problem as one-dimensional by taking the field operators to be of the form,
\begin{equation}
\hat{\psi}_\sigma({\mathbf{x}})=\phi(\rho)\,\hat{\psi}_\sigma(x),
\end{equation}
Carrying out the $\rho$ integration in eq.~\ref{eq:6.1}, we determine the effective one-dimensional interaction Hamiltonian,
\begin{equation}
\hat{H}'=\tilde{g}\int dx\,\hat{\psi}_\uparrow^\dagger(x)\hat{\psi}_\downarrow^\dagger(x) H'\hat{\psi}_\downarrow(x)\hat{\psi}_\uparrow(x).
\label{eq:6.5}
\end{equation}
where $\tilde{g}\equiv g\,\bar{n}_\perp$ and
\begin{equation}
\bar{n}_\perp=\int 2\pi\rho d\rho\,[n_\perp(\rho)]^2.
\label{eq:nperpav}
\end{equation}
Here, we have let  $|\phi(\rho)|^2\rightarrow n_\perp(\rho)$, where $\int 2\pi \rho d\rho\,n_\perp(\rho)=1$. Eq.~\ref{eq:nperpav} determines an effective mean transverse density, $\bar{n}_\perp$, as a fraction per unit transverse area. Using eq.~\ref{eq:3.5}, eq.~\ref{eq:6.5} takes the form
\begin{eqnarray}
\hat{H}'&=&\tilde{g}\!\!\!\!\sum_{E_1,E_2,E_1',E_2'}\int dx\,\phi^*_{E_1'}(x)\phi^*_{E_2'}(x) \phi_{E_2}(x)\phi_{E_1}(x)\nonumber\\
& &\hspace{0.25in}\times\, \hat{a}^\dagger_\uparrow(E_1')\hat{a}^\dagger_\downarrow(E_2')\hat{a}_\downarrow(E_2)\hat{a}_\uparrow(E_1).
\label{eq:6.6}
\end{eqnarray}
With the anticommutation relations, eq.~\ref{eq:2.2}, we can rewrite the operator product of eq.~\ref{eq:6.6} as
\begin{equation}
\hat{O}'\equiv \hat{a}^\dagger_\uparrow(E_1')\hat{a}_\uparrow(E_1)\hat{a}^\dagger_\downarrow(E_2')\hat{a}_\downarrow(E_2).
\label{eq:6.7}
\end{equation}

We simplify the interaction Hamiltonian by using a mean field approximation to evaluate eq.~\ref{eq:6.7}.
To first order, we obtain
\begin{eqnarray}
\hat{O}'&\simeq&\langle \hat{a}^\dagger_\uparrow(E_1')\,\hat{a}_\uparrow(E_1)\rangle\,\hat{a}^\dagger_\downarrow(E_2')\,\hat{a}_\downarrow(E_2)\nonumber\\
& &+\langle\, \hat{a}^\dagger_\downarrow(E_2')\,\hat{a}_\downarrow(E_2)\rangle\,\hat{a}^\dagger_\uparrow(E_1')\,\hat{a}_\uparrow(E_1)\nonumber\\
& &-\langle\, \hat{a}^\dagger_\uparrow(E_1')\,\hat{a}_\downarrow(E_2)\rangle\,\hat{a}^\dagger_\downarrow(E_2')\,\hat{a}_\uparrow(E_1)\nonumber\\
& &-\langle \hat{a}^\dagger_\downarrow(E_2')\,\hat{a}_\uparrow(E_1)\rangle\,\hat{a}^\dagger_\uparrow(E_1')\,\hat{a}_\downarrow(E_2),
\label{eq:21.4}
\end{eqnarray}
where $\langle ...\rangle$ denotes a thermal average, which vanishes unless the energy arguments are the same. Further,
we will require a thermal average of the Heisenberg equations of motion, i.e., $\langle[\hat{O}',\hat{S}_i(E)]\rangle$. This will vanish unless the energy arguments in the operator factors are the same. Hence,
Eq.~\ref{eq:6.6} can be rewritten as
\begin{eqnarray}
\hat{H}'&=&\tilde{g}\sum_{\tilde{E},E'}\int dx\,|\phi_{E'}(x)|^2|\phi_{\tilde{E}}(x)|^2\nonumber\\
& &\times\Big\{\langle \hat{a}^\dagger_\uparrow(E')\,\hat{a}_\uparrow(E')\rangle\,\hat{a}^\dagger_\downarrow(\tilde{E})\,\hat{a}_\downarrow(\tilde{E})\nonumber\\
& &+\langle \hat{a}^\dagger_\downarrow(E')\,\hat{a}_\downarrow(E')\rangle\,\hat{a}^\dagger_\uparrow(\tilde{E})\,\hat{a}_\uparrow(\tilde{E})\nonumber\\
& &-\langle \hat{a}^\dagger_\uparrow(E')\,\hat{a}_\downarrow(E')\rangle\,\hat{a}^\dagger_\downarrow(\tilde{E})\,\hat{a}_\uparrow(\tilde{E})\nonumber\\
& &-\langle \hat{a}^\dagger_\downarrow(E')\,\hat{a}_\uparrow(E')\rangle\,\hat{a}^\dagger_\uparrow(\tilde{E})\,\hat{a}_\downarrow(\tilde{E})\Big\}.
\label{eq:21.5}
\end{eqnarray}
With the collective spin operators, eq.~\ref{eq:3.4}, we rewrite eq.~\ref{eq:21.5} as
\begin{eqnarray}
\hat{H}'&=&2\,\tilde{g}\sum_{\tilde{E},E'}\int dx\,|\phi_{E'}(x)|^2|\phi_{\tilde{E}}(x)|^2\nonumber\\
& &\times\Big\{\frac{1}{4}N(E')\,\hat{N}(\tilde{E})-\,{\mathbf{S}}(E')\cdot\hat{\mathbf{S}}(\tilde{E})\Big\},
\label{eq:22.3}
\end{eqnarray}
where $\hat{N}(\tilde{E})=\hat{N}_\uparrow(\tilde{E})+\hat{N}_\downarrow(\tilde{E})$ is the total number operator and $\hat{\mathbf{S}}(\tilde{E})$ is the total spin vector operator for atoms of energy $\tilde{E}$.  $N(E')$ is a c-number scalar and ${\mathbf{S}}(E')$ is a c-number vector, i.e., the corresponding thermal averaged Heisenberg operators for energy $E'$.

To evaluate of the collisional contribution to the Heisenberg equations of motion, we require $[\hat{H}',\hat{\mathbf{S}}(E)]$. Here,
$[\hat{N}(\tilde{E}),\hat{\mathbf{S}}(E)]=0$, and using eq.~\ref{eq:Scomm}, $[{\mathbf{S}}(E')\cdot\hat{\mathbf{S}}(\tilde{E}),\hat{{\mathbf S}}(E)]=-i\,{\mathbf{S}}(E')\times\hat{\mathbf{S}}(\tilde{E})\,\delta_{\tilde{E},E}$. With eq.~\ref{eq:5.5}, the Heisenberg equation $\dot{\hat{{\mathbf S}}}(E,t)=\frac{i}{\hbar}\left[\hat{H}_0+\hat{H}',\hat{\mathbf{S}}(E,t)\right]$ for the spin vector operator of energy $E$  takes the simple form,
\begin{eqnarray}
\frac{\partial\hat{\mathbf{S}}(E,t)}{\partial t}&=&{\mathbf{\Omega}}(E,t)\times\hat{\mathbf{S}}(E,t)\nonumber\\
& &\hspace{-0.25in}+\,\sum_{E'}g(E',E)\,{\mathbf{S}}(E',t)\times\hat{\mathbf{S}}(E,t).
\label{eq:22.4}
\end{eqnarray}
In eq.~\ref{eq:22.4},
\begin{equation}
g(E',E)=-\frac{2\,g\,\bar{n}_\perp}{\hbar}\,I(E',E),
\label{eq:15.3}
\end{equation}
where $I(E',E)\equiv\int dx\,|\phi_{E'}(x)|^2|\phi_{E}(x)|^2$ with $\bar{n}_\perp$ is given by eq.~\ref{eq:nperpav}, and $g=4\pi\hbar^2\,a_S/m$.

In our experiments, where the energy $E>>\hbar\bar{\omega}_x$,  $|\phi_{E}(x)|^2$ can be evaluated in a WKB approximation, neglecting the rapid spatial oscillation from the phase,
\begin{equation}
|\phi_E(x)|^2\simeq\frac{\Theta[a(E)-|x|]}{\pi\sqrt{a^2(E)-x^2}},
\label{eq:15.8}
\end{equation}
where $a(E)=\sqrt{2E/(m\bar{\omega}^2_x)}$ is the classical turning point and $\Theta$ is a Heaviside function.
Then, the $x$-integral in eq.~\ref{eq:15.3} takes the form
$$I(E',E)=\frac{1}{\pi^2a_{min}}\int_{-1}^1\frac{du}{\sqrt{\left[\frac{E}{E_{min}}-u^2\right]\left[\frac{E'}{E_{min}}-u^2\right]}}, $$
where we have taken $x=u\,a_{min}$. Here, $a_{min}=\sqrt{2E_{min}/(m\bar{\omega}^2_x)}$ determines the overlap region, with $E_{min}$ the minimum of $E,E'$. Using $u=\sin\theta$, and by considering separately the cases $E_{min}=E<E'$ and $E_{min}=E'<E$, we obtain
$$I(E',E)=\frac{1}{\pi^2}\sqrt{\frac{m\bar{\omega}^2_x}{2|E-E'|}}\int_{-\pi/2}^{\pi/2}\frac{d\theta}{\sqrt{1+\frac{E_{min}}{|E-E'|}\cos^2\theta}}.$$
The integral is readily evaluated, yielding
\begin{eqnarray}
g(E',E)&=&-\frac{4\,g\,\bar{n}_\perp}{\pi^2\hbar}\,\sqrt{\frac{m\bar{\omega}^2_x}{2|E-E'|}}\nonumber\\
& &\times\,{\rm EllipticK}\left[-\frac{E_{min}}{|E-E'|}\right],
\label{eq:g}
\end{eqnarray}
where $E'\neq E$, since the sum in the last term of eq.~\ref{eq:22.4} vanishes for $E'=E$, i.e., we can take $g(E'=E,E)=0$ in eq.~\ref{eq:22.4}.

Taking the thermal average of the evolution equations, we replace the vector operators by the c-number vectors ${\mathbf{S}}(E,t)\equiv\langle\hat{\mathbf{S}}(E,t)\rangle$. Since $E>>\hbar\bar{\omega}_x$, we  evaluate eq.~\ref{eq:22.4} in the continuum limit. We replace the sum $\sum_E'\equiv\sum_n'$ by $\int\frac{dE'}{\hbar\bar{\omega}_x}$ and define
\begin{equation}
\frac{{\mathbf{S}}(E,t)}{\hbar\bar{\omega}_x}\equiv\frac{N}{2}\tilde{\mathbf{S}}(E,t),
\label{eq:stilde}
\end{equation}
where $N=N_\uparrow+N_\downarrow$ is the total number of atoms. Then,
\begin{eqnarray}
\frac{\partial\tilde{\mathbf{S}}(E,t)}{\partial t}&=&{\mathbf{\Omega}}(E,t)\times\tilde{\mathbf{S}}(E,t)\label{eq:19.6}\\
& &\hspace{-0.25in}+\,\int dE'\,\tilde{g}(E',E)\,\tilde{\mathbf{S}}(E',t)\times\tilde{\mathbf{S}}(E,t),\nonumber
\end{eqnarray}
where $\tilde{g}(E',E)\equiv \frac{N}{2}\,g(E',E)$ has a dimension of $s^{-1}$.  Note that the factor $N/2$ in eq.~\ref{eq:stilde} is defined to be consistent with the spin operators of eq.~\ref{eq:3.4}, i.e., with all atoms in the ground $\uparrow$ hyperfine state, the total spin in the z-direction is $N/2$.

The integral term in eq.~\ref{eq:19.6} conserves the total spin vector $\int dE\,\tilde{\mathbf{S}}(E,t)$, since $\tilde{g}(E,E')$ is symmetric under $E'\leftrightarrow E$ and the cross product is antisymmetric. In contrast,  ${\mathbf{\Omega}}(E)$ is an energy dependent rotation rate that does not conserve the total spin $\tilde{\mathbf{S}}(E,t)$. However, without radio frequency excitation, ${\mathbf{\Omega}}(E)$ is along the $z$-axis and the z-component of the total spin $\int dE\,\tilde{S}_z(E)$ is conserved. Finally, since eq.~\ref{eq:19.6} describes a rotation of $\tilde{\mathbf{S}}(E,t)$,  $|\tilde{\mathbf{S}}(E,t)|\equiv S(E)$ is conserved for each $E$.

We integrate eq.~\ref{eq:19.6}  subject to the initial condition that all atoms are in the lower hyperfine (spin-up) state. A radio frequency pulse is then used to prepare a collective spin vector with components in the $x-y$ plane. With eq.~\ref{eq:stilde}, the thermal averaged z-component of the initial collective spin operator, eq.~\ref{eq:3.4}, requires
\begin{equation}
\tilde{S}_z(E,t=0)=S(E)=P(E),
\label{eq:18.4}
\end{equation}
where  $P(E,T)$ is the fraction of atoms with axial energy $E$ at temperature $T$ and $\int_0^\infty dE\, P(E)=1$ in the continuum limit. In the high temperature limit,
\begin{equation}
P(E)=\frac{1}{Z}\,e^{-\frac{E}{k_BT}},
\label{eq:18.3}
\end{equation}
with the partition function $Z=\int_0^\infty dE e^{-\frac{E}{k_BT}}=k_BT$. In the low temperature limit, $T\rightarrow 0$, we use the occupation number for a Fermi distribution in three dimensions and sum over the energies in the two perpendicular directions to obtain the normalized axial ($x$) energy distribution,
\begin{equation}
P(E)=\frac{3}{E_F}\left(1-\frac{E}{E_F}\right)^2\,\Theta\left(1-\frac{E}{E_F}\right),
\label{eq:5.2}
\end{equation}
where for $N_\uparrow=N$, $E_F=(6N)^{1/3}\hbar\bar{\omega}$, with $\bar{\omega}\equiv(\omega_\perp^2\bar{\omega}_x)^{1/3}$.

The measured axial spin density  profiles are given by the continuum limit of eq.~\ref{eq:3.7},
\begin{equation}
{\mathbf{S}}(x,t)=\frac{N}{2}\int dE\,|\phi_E(x)|^2\,\tilde{\mathbf{S}}(E,t),
\label{eq:spindensity}
\end{equation}
where we neglect coherence between states of different energy and $\int dx\,{\mathbf{S}}(x,t)=\frac{N}{2}\int dE\,\tilde{\mathbf{S}}(E,t)$.
Evaluation of eq.~\ref{eq:spindensity} is simplified by rewriting the WKB wave functions of eq.~\ref{eq:15.8} in the form
\begin{equation}
|\phi_E(x)|^2=\frac{\bar{\omega}_x}{\pi}\int_0^\infty dp_x\,\delta\left(E-\frac{p_x^2}{2m}-\frac{m\bar{\omega}_x^2}{2}x^2\right)
\label{eq:3.2}
\end{equation}
so that the spin density is
\begin{equation}
{\mathbf{S}}(x,t)=\frac{N}{2}\frac{\bar{\omega}_x}{\pi}\int_0^\infty dp_x\,\tilde{\mathbf{S}}\left(\frac{p_x^2}{2m}+\frac{m\bar{\omega}_x^2}{2}x^2,t\right).
\label{eq:spindensity1}
\end{equation}

The initial spatial densities for the spin components are similarly determined. For the degenerate gas, we approximate the energy distribution by the zero temperature limit, eq.~\ref{eq:5.2}, as discussed above. The corresponding spatial density for each spin component, just after preparation, is then a normalized zero temperature Thomas-Fermi profile. Analogous to eq.~\ref{eq:spindensity}, using eq.~\ref{eq:15.8} (or eq.~\ref{eq:3.2}), it is easy to show that the initial density profiles for each state are of the one dimensional Thomas-Fermi form,
\begin{eqnarray}
n_{\uparrow,\downarrow}(x,0)&=&N_{\uparrow,\downarrow}\int dE\,|\phi_E(x)|^2\,P(E)\label{eq:TFdensity}\\
& &\hspace{-0.5in}=N_{\uparrow,\downarrow}\,\frac{16}{5\pi\,\sigma_{Fx}}\left(1-\frac{x^2}{\sigma_{Fx}^2}\right)^{5/2}
\Theta\left(1-\frac{x^2}{\sigma_{Fx}^2}\right),\nonumber
\end{eqnarray}
where $\sigma_{Fx}=\sqrt{2E_F/(m\bar{\omega}_x^2)}$ is the Fermi radius and $N_\uparrow=N_\downarrow=N/2$ for a balanced mixture. As the energy distribution for the atoms does not change in time, the spatial profile for the total density $n(x)$ is time independent, i.e.,  $n_\uparrow(x,t)+n_\downarrow(x,t)=n_\uparrow(x,0)+n_\downarrow(x,0)=n(x)$, as shown in Fig.~1 of the main paper. For the non-degenerate gas, the Maxwell-Boltzmann energy distribution of eq.~\ref{eq:18.3} yields the corresponding gaussian spatial profile.

\subsection{Small Angle Approximation}
We can make contact with the first order, large Dicke gap approximation of Koller et al~\cite{KollerReySpinDep}, by considering the evolution equations for small amplitude spin waves, expressed in terms of angles.
As the magnitude of $|\tilde{\mathbf{S}}(E,t)|\equiv S(E)$ is conserved for each $E$, where $\int dE\,S(E)=1$, we can write the spin components in terms of two angles, a polar angle $\theta_E$ and an azimuthal angle, $\varphi_E$,
\begin{eqnarray}
\tilde{S}_x(E,t)&=&S(E)\,\sin\theta_E(t)\cos\varphi_E(t)\nonumber\\
\tilde{S}_y(E,t)&=&S(E)\,\sin\theta_E(t)\sin\varphi_E(t)\nonumber\\
\tilde{S}_z(E,t)&=&S(E)\,\cos\theta_E(t).
\label{eq:angle1.2}
\end{eqnarray}
Using eq.~\ref{eq:19.6}, it is straightforward to obtain the evolution equations for the angles. For times after the radio-frequency preparation pulse, the $\dot{\tilde{S}}_z$ equation yields
\begin{equation}
\dot{\theta}_E=\int dE'\,\tilde{g}(E,E')\,S(E')\,\sin\theta_{E'}\,\sin(\varphi_{E'}-\varphi_{E})
\label{eq:angle1.3}
\end{equation}
and $\cos\varphi_E\,\dot{\tilde{S}}_y-\sin\varphi_E\,\dot{\tilde{S}}_x$ gives
\begin{eqnarray}
\dot{\varphi}_E&=&\gamma\,E+\int dE_1'\,\tilde{g}(E,E_1')\,S(E_1')\label{eq:angle1.4}\\
& &\hspace{-0.25in}\times[\,\cos\theta_{E_1'}-\cot\theta_E\,\sin\theta_{E_1'}\,\cos(\varphi_E-\varphi_{E_1'})],\nonumber
\end{eqnarray}
where $\gamma\,E$ is the energy-dependent rotation rate about the z-axis,  eq.~\ref{eq:1.7}, i.e., $\Omega(E)=-\delta\omega_x/(\hbar\bar{\omega}_x)\,E\equiv\gamma\,E$. Here, we take the initial conditions to be $\tilde{S}_x(E,t=0)=S(E)$ and $\tilde{S}_z(E,t=0)=\tilde{S}_y(E,t=0)=0$,  just after the radio frequency pulse.
From eq.~\ref{eq:angle1.3}, we see that $\int dE\,\dot{\tilde{S}}_z(E,t)=-\int dE\,S(E)\,\sin\theta_E\,\dot{\theta}_E=0$, since $\sin(\varphi_{E'}-\varphi_{E})$ is odd in $E',E$ and $\int dE\,\tilde{S}_z(E,t)$ is conserved as it should be.

The angle equations take a simple approximate form for small amplitude spin waves, where $\theta_E=\pi/2+\delta\theta_E$ with  $\delta\theta_E<<1$. Then,
\begin{equation}
\tilde{S}_z(E,t)\simeq -S(E)\,\delta\theta_E(t)
\label{eq:angle1.8}
\end{equation}
and the spatial profile, eq.~\ref{eq:spindensity}, is given by
\begin{equation}
S_z(x,t)=-\frac{N}{2}\,\int dE\,|\phi_E(x)|^2\,S(E)\,\delta\theta_E(t),
\label{eq:angle3.8}
\end{equation}
where $|\phi_E(x)|^2$ is easily evaluated using the WKB approximation.

For $\gamma\, E>>\tilde{g}(E,E')$, with $\sin\theta_{E'}\simeq 1$ and $\varphi_{E'}-\varphi_{E}\simeq \gamma(E'-E)t$, eq.~\ref{eq:angle1.3} immediately yields
\begin{eqnarray}
\delta\theta_E(t)&\simeq&\int dE'\,\tilde{g}(E,E')\,S(E')\nonumber\\
& &\times\,\frac{1-\cos[\gamma(E'-E)t]}{\gamma(E'-E)}.
\end{eqnarray}

To make contact with the first order, large Dicke gap approximation of Koller et al~\cite{KollerReySpinDep}, we consider the opposite limit, $\tilde{g}(E,E') >>\gamma\, E$. Here, we make the simplifying assumption that $\tilde{g}(E,E')\simeq\bar{\Omega}_g$ is energy independent.
Then we can approximate $\varphi_E-\varphi_{E'} <<1$, over the relevant time scale $t\simeq 1/\bar{\Omega}_g$ and
eqs.~\ref{eq:angle1.3}~and~\ref{eq:angle1.4} take the simple forms,
\begin{eqnarray}
\delta\dot{\theta}_E&=&\bar{\Omega}_g\,\int dE'\,S(E')\,(\varphi_{E'}-\varphi_E)\label{eq:angle2.2}\\
\dot{\varphi}_E&=&\gamma\,E+\bar{\Omega}_g\,\int dE_1'\,S(E_1')\,(\delta\theta_E-\delta\theta_{E_1'}).\nonumber
\end{eqnarray}
Differentiating the first equation with respect to $t$ yields,
\begin{equation}
\delta \ddot{\theta}_E=\bar{\Omega}_g\,\int dE'\,S(E')\,(\dot{\varphi}_{E'}-\dot{\varphi}_E).
\label{eq:angle2.4}
\end{equation}
From the second equation,
\begin{equation}
\dot{\varphi}_{E'}-\dot{\varphi}_E=\gamma\,(E'-E)+\bar{\Omega}_g(\delta\theta_{E'}-\delta\theta_E),
\label{eq:angle2.6}
\end{equation}
where we have used $\int dE_1'\,S(E_1')=1$. After substituting eq.~\ref{eq:angle2.6} into eq.~\ref{eq:angle2.4}, we take $\int dE'\,S(E')\,\delta\theta_{E'}=0$. Here, we assume for simplicity that the initial spin is in the x-y plane, so that the conserved total $\tilde{S}_z$ vanishes. Then,
\begin{equation}
\delta\ddot{\theta}_E+\bar{\Omega}_g^2\,\delta\theta_E=\bar{\Omega}_g\gamma(\bar{E}-E),
\end{equation}
where $\bar{E}\equiv\int dE'\,S(E')\,E'$.  For the initial conditions, $\delta\theta_E(0)=0$ and $\delta\dot{\theta}_E(0)=0$,
\begin{equation}
\delta\theta_E(t)=\frac{\gamma\,(\bar{E}-E)}{\bar{\Omega}_g}\,[1-\cos(\bar{\Omega}_g t)].
\label{eq:angle3.7}
\end{equation}
With eq.~\ref{eq:angle3.8}, we see that eq.~\ref{eq:angle3.7} is equivalent to eq. 3 of Koller et al~\cite{KollerReySpinDep}, which was obtained by first order perturbation theory in the Dicke spin state basis.

\subsection{Numerical Implementation}
\label{sec:numerical}

To determine $\tilde{\mathbf{S}}(E,t)$ from eq.~\ref{eq:19.6}, we divide the energy range into discrete intervals $\Delta E$, taking  $E=(n-1)\Delta E$, with $n$ an integer, $1\leq n\leq n_{\rm max}$. Typically, $n_{max}=500$. This method determines the spin components $i=x,y,z$ as column vectors in discrete energy space,  $\tilde{S}_i^{\rm \,discr}(n,t)$, where $n$ labels the row (rather than the harmonic oscillator state). We take $\tilde{\mathbf{S}}(E,t)=\tilde{\mathbf{S}}^{\rm \,discr}(n,t)/\Delta E$ in eq.~\ref{eq:19.6}.
With the replacement $\int dE'/\Delta E= \int dn'\rightarrow\sum_{n'}$, the discrete energy evolution equations  are
\begin{eqnarray}
\frac{\partial\tilde{\mathbf{S}}^{\rm \,discr}(n,t)}{\partial t}&=&{\mathbf{\Omega}}(n,t)\times\tilde{\mathbf{S}}^{\rm \,discr}(n,t)\label{eq:3.6b}\\
& &\hspace{-0.5in}+\,\sum_{n'}\tilde{g}(n',n)\,\tilde{\mathbf{S}}^{\rm \,discr}(n',t)\times\tilde{\mathbf{S}}^{\rm \,discr}(n,t).\nonumber
\end{eqnarray}
where
\begin{equation}
\tilde{g}(n',n)=\frac{\tilde{\Omega}}{\sqrt{|n-n'|}}\,\,{\rm EllipticK}\left[-\frac{n_{min}-1}{|n-n'|}\right].
\label{eq:tildeg}
\end{equation}
Here, $n_{min}$ is the minimum of $n$ and $n'$ and
\begin{equation}
\tilde{\Omega}=-\frac{N}{2}\frac{4\,g\,\bar{n}_\perp}{\pi^2\hbar}\,\sqrt{\frac{m\bar{\omega}^2_x}{2\Delta E}},
\label{eq:Omegag1}
\end{equation}
with $g=4\pi\hbar^2\,a_S/m$.

We define $\Delta E$ differently for the high and low temperature limits. In the low temperature limit,
we take $\Delta E=s\,E_F$. Since $0\leq E\leq E_F$, we have $s=1/(n_{\rm max}-1)$.
In the high temperature limit, we choose $\Delta E=s\,k_B T$  and take $s$ so that  $\exp[-s\,(n_{\rm max}-1)]$ is negligible. For both cases, it is convenient to let $\Delta E=s\,\frac{1}{2}m\bar{\omega}_x^2\sigma_x^2$. Then, for $T=0$,   $\sigma_x=\sqrt{2E_F/(m\bar{\omega}_x^2)}\,\equiv\sigma_{Fx}$ is the Fermi radius, which is measured in the experiments. For the high temperature limit, $\sigma_x=\sqrt{2k_BT/(m\bar{\omega}_x^2)}$  is the measured gaussian (Boltzmann factor) $1/e$ radius.
With $\Delta E=s\,\frac{1}{2}m\bar{\omega}_x^2\sigma_x^2$, eq.~\ref{eq:Omegag1} yields
\begin{equation}
\tilde{\Omega}=-\frac{1}{\sqrt{s}}\frac{4\,h}{\pi^2}\frac{a_S}{m}\frac{{\bar{n}_\perp}N}{\sigma_x}\equiv-\frac{1}{\sqrt{s}}\,\Omega_{MF},
\label{eq:Omegag}
\end{equation}
where we have defined the mean field frequency $\Omega_{MF}$, $h=2\pi\hbar$, and $\bar{n}_\perp$ given by eq.~\ref{eq:nperpav}. In the low temperature limit, with $n_\perp(\rho)=3(1-\rho^2/\sigma_{F\perp}^2)/(\pi\sigma_{F\perp}^2)$, we obtain $\bar{n}_\perp=\frac{9}{5\pi\sigma_{F\perp}^2}$. In the high temperature limit, with $n_\perp(\rho)=\exp[-\rho^2/\sigma_\perp^2]/(\pi\sigma_\perp^2)$, we obtain $\bar{n}_\perp=\frac{1}{2\pi\sigma_{\perp}^2}$. Then,
\begin{eqnarray}
\Omega_{MF}&=&\frac{9}{20\pi}\frac{2h\,a_S}{m}\,n_{F0}\quad\,\mbox{\rm $T=0$}\nonumber\\
\Omega_{MF}&=&\frac{1}{\pi^{3/2}}\frac{2h\,a_S}{m}\,n_{0}\quad\mbox{\rm \hspace{0.05in}High $T$}.
\label{eq:meanFfreq}
\end{eqnarray}
Here $n_{F0}=8\,N/(\pi^2\sigma_{F\perp}^2\sigma_{Fx})$ is the 3D central density for a $T=0$ Thomas-Fermi profile with $\sigma_{F\perp}=\sqrt{2E_F/(m\omega_\perp^2)}$ and $n_0=N/(\pi^{3/2}\sigma_\perp^2\sigma_x)$ is the 3D central density in the Boltzmann limit, where $\sigma_\perp=\sqrt{2k_BT/(m\omega_\perp^2)}$\,.

With our choices of $\Delta E$, the initial conditions are analogous to eq.~\ref{eq:18.4},
\begin{equation}
\tilde{S}^{\rm \,discr}_z(n,t=0)=P(n),
\label{eq:5.3}
\end{equation}
where $P(n)=\tilde{p}(n)/\tilde{Z}$ with $\tilde{Z}=\sum_{n=1}^{n_{\rm max}}\tilde{p}(n)$. In the high temperature limit, $\tilde{p}(n)=s\,\exp[-s(n-1)]$,  and for the $T=0$ limit, $\tilde{p}(n)=3s\,[1-s(n-1)]^2$. Note that $\tilde{Z}\simeq 1$ for large $n_{\rm max}$.

Now we evaluate the first term on the right side of eq.~\ref{eq:3.6b}, which is the energy-dependent frequency ${\mathbf{\Omega}}(n,t)=\hat{e}_z\,\Omega_z(n)+{\mathbf{\Omega}}_{\rm Rabi}(t)$. As discussed above,  $\Omega_z(n)$ arises from the bias magnetic field curvature. For a general radio-frequency excitation with a time-dependent detuning $\Delta(t)$ and Rabi frequency $\Omega_R(t)$, ${\mathbf{\Omega}}_{\rm Rabi}(t)=\hat{e}_z\,\Delta(t)+\hat{e}_y\Omega_R(t)$. Using $E=(n-1)s\,E_F$ for the $T=0$ limit and $E=(n-1)s\,k_BT$ in the high temperature limit, we have
\begin{equation}
\Omega_z(n)\equiv(n-1)\,\Omega_{z0}.
\label{eq:1.7n}
\end{equation}
where $\Omega_{z0}=-\delta\omega_x\, s\, E_F/(\hbar\bar{\omega}_x)$ at $T=0$ and  $\Omega_{z0}=-\delta\omega_x\, s\, k_BT/(\hbar\bar{\omega}_x)$ in the high temperature limit.

Next, we evaluate the resonance frequency difference,  $\delta\omega_x=\omega_{x\downarrow}-\omega_{x\uparrow}$, which arises from the curvature of the bias magnetic field in the axial $x$ direction, $\Delta B_z=x^2\,B_z''(0)/2$. The harmonic oscillation frequencies for the upper hyperfine state ($\downarrow$) and lower hyperfine state ($\uparrow$)  are determined by the sum of optical and magnetic spring constants,
\begin{equation}
\omega^2_{x\downarrow,\uparrow}=\omega^2_{\rm opt}+\omega^2_{{\rm mag}\downarrow,\uparrow}=\omega^2_{\rm opt}+\frac{1}{m}\frac{\partial^2B_z}{\partial^2x}\frac{\partial E_{\downarrow,\uparrow}}{\partial B},
\label{eq:1.2}
\end{equation}
where $\omega_{\rm opt}$ arises from the optical trap and $\omega_{\rm mag}$ from the bias field curvature.

For our experiments in $^6$Li, the hyperfine energies $E_{\downarrow,\uparrow}$ are dominated by the Zeeman shift of the (spin down) electron for each of the lowest three hyperfine states, while the much smaller difference $E_\downarrow-E_\uparrow$ arises from the difference between the nuclear parts of the magnetic moment and the difference in the hyperfine mixing. Then, with $\omega_{\rm mag}^2\equiv(\omega_{\rm mag\downarrow}^2+\omega_{\rm mag\uparrow}^2)/2$ and $\bar{\omega}_x^2\equiv\omega^2_{\rm opt}+\omega_{\rm mag}^2$, we have
\begin{eqnarray}
\omega_{x\downarrow,\uparrow}&=&\sqrt{\omega^2_{\rm opt}+\omega_{\rm mag}^2\pm\frac{\omega^2_{{\rm mag}\downarrow}-\omega^2_{{\rm mag}\uparrow}}{2}}\nonumber\\
&\simeq&\bar{\omega}_x\left(1\pm\frac{\omega^2_{{\rm mag}\downarrow}-\omega^2_{{\rm mag}\uparrow}}{4\,\bar{\omega}_x^2}\right)
\label{eq:1.4}
\end{eqnarray}
and
$$\frac{\delta\omega_x}{\bar{\omega}_x}=\frac{\omega_{x\downarrow}-\omega_{x\uparrow}}{\bar{\omega}_x}=\frac{\omega_{\rm mag}^2}{\bar{\omega}_x^2}\frac{\omega^2_{{\rm mag}\downarrow}-\omega^2_{{\rm mag}\uparrow}}{2\,\omega_{\rm mag}^2}. $$
Then,
\begin{equation}
\delta\omega_x=\frac{\omega_{\rm mag}^2}{\bar{\omega}_x}\left(\frac{\frac{\partial E_{\downarrow}}{\partial B}-\frac{\partial E_{\uparrow}}{\partial B}}{\frac{\partial E_{\downarrow}}{\partial B}+\frac{\partial E_{\uparrow}}{\partial B}}\right)\simeq\frac{\omega_{\rm mag}^2}{\bar{\omega}_x}\frac{\hbar\omega'_{\downarrow\uparrow}}{g_J\mu_B},
\label{eq:1.6}
\end{equation}
where $\omega'_{\downarrow\uparrow}$ is the tuning rate of the transition, with $\downarrow$ the {\it upper} hyperfine state.
Here, we have assumed that the denominator of eq.~\ref{eq:1.6} is approximately twice the Zeeman tuning rate of a spin-down electron, $2\times g_J\mu_B/2=-2\pi\times2.8$ MHz/G, as is the case for our experiments near the zero crossings of $^6$Li. For our experiments, $\omega_{\rm mag}^2=(2\pi\times 20.5\,{\rm Hz})^2\,{\rm B(G)}/834$. For the degenerate gas, $\bar{\omega}_x=2\pi\times 23$ Hz, $\omega_\perp=2\pi\times 625$ Hz; for the high temperature gas, $\bar{\omega}_x=2\pi\times 174$ Hz, $\omega_\perp=2\pi\times 5.77$ kHz.

For a mixture of two hyperfine states, as noted above, $\downarrow$ denotes the upper hyperfine state, and $\uparrow$ denotes the lower hyperfine state. The hyperfine energies for the three lowest states of $^6$Li, denoted $1,2,3$ in order of increasing energy, yield the tuning rates which appear in the numerator of eq.~\ref{eq:1.6}: $\omega_{21}'[527\,{\rm G}]=2\pi\times 3.61$ kHz/G and $\omega_{32}'[589\,{\rm G}]=-2\pi\times 12.3$ kHz/G, $\omega_{31}'[568\, {\rm G}]=-2\pi\times 10.3$ kHz/G. With $\bar{\omega}_x=2\pi\times 23$ Hz, we obtain $\delta\omega_x=-2\pi\times 14.9$ mHz  for a $1-2$ mixture near $527$ G, $\delta\omega_x=+2\pi\times 56.7$ mHz for a $2-3$ mixture near $589$ G, and $\delta\omega_x=+2\pi\times 45.8$ mHz for a $1-3$ mixture near $568$ G.

Numerical evaluation of eq.~\ref{eq:3.6b} yields the tables $\{n-1,\tilde{S}_i^{\rm \,discr}(n,t)\}$ for $1\leq n\leq n_{\rm max}$. Note that $n-1$ is used as the independent variable so that $E=(n-1)\Delta E=0$ for $n=1$. The energy-dependent $\tilde{\mathbf{S}}^{\rm \,discr}(n,t)$ is then converted to an interpolator function of $(n-1)=E/\Delta E$ and eq.~\ref{eq:spindensity1} used to find the spin density $\mathbf{S}(x,t)$.

\subsection{Energy Dependent Scattering Length}
For experiments in the non-degenerate regime at higher temperatures, we find that the energy dependence of the scattering length cannot be neglected. This energy dependence strongly modifies the spin-density profiles for small positive scattering lengths, as shown in Fig.~6 of the main paper, and produces a shift of the zero crossing field.  We include this dependence in $g(E',E)$ of eq.~\ref{eq:15.3} by replacing the energy-independent s-wave scattering length $a_S$ with an energy dependent scattering length $a(E',E)$. The s-wave scattering length is given by the energy-dependent scattering amplitude $f(k)$,
\begin{equation}
a[B,k]=f\left(-2\mu_B B+\frac{\hbar^2{\mathbf k}^2}{2\mu}\right),
\label{eq:1.1a}
\end{equation}
where $\hbar\,{\mathbf k}$ is the relative momentum and $\mu = m/2$ is the reduced mass. The applied bias magnetic field $B_z\equiv B$ tunes the energy of a colliding pair in the triplet channel downward, at a rate $-2\mu_B B $, with $\mu_B$ the Bohr magneton.  For our experiments in the degenerate regime, where the relative kinetic energy term in eq.~\ref{eq:1.1a} is negligible, we assume that the scattering length  varies linearly with applied magnetic field near the zero crossing field $B_0$,
\begin{equation}
a(B)=a'\,(B-B_0),
\label{eq:scattlength}
\end{equation}
where the tuning rate of the scattering length $a'$ is given in the main text in units of $a_0/$G, where $a_0$ is the Bohr radius.

Including the relative kinetic energy $K_{\rm rel}$ in eq.~\ref{eq:1.1a} is equivalent to replacing the magnetic field $B$ by an effective magnetic field,
\begin{equation}
B_{eff}=\langle B_z\rangle -\frac{\langle K_{\rm rel}\rangle}{2\mu_B}.
\label{eq:1.5a}
\end{equation}
Here, we include an additional average of the spatially varying bias field $B_z$ over the position of the center of mass (CM) of a colliding atom pair.

We begin by evaluating $\langle B_z\rangle$. The bias field is cylindrically symmetric about the $z$ axis, and oriented perpendicular to the long $x$-axis of the trapped cloud, so that $B_z=B_{z0}[1+b(z^2-(x^2+y^2)/2)]$, where $B_{z0}$ is the bias field at the cloud center and $b\,B_{z0}$ is the field curvature. For the cigar-shaped clouds utilized in the experiments,  the variation of $B_z$ in the $z$ and $y$ directions is negligible compared to that in the $x$ direction, so that $B_z(x)=B_{z0}[1-b\, x^2/2]$. We determine $b\,B_{z0}$ from the measured spring constant  of the resulting harmonic confining potential, $-\mu_B B_z(x)$, where for $^6$Li, the magnetic moment, $+\mu_B$, of the three lowest hyperfine states at high $B$ field is dominated by the electron spin down contribution, $m_s=-1/2$. With $\mu_B\,b\,B_{z0}\equiv m\omega_{\rm mag}^2$, where $\omega_{\rm mag}$ is given in \S~\ref{sec:numerical}, the bias field, averaged over the center of mass position, is then $\langle B_z\rangle=B_{z0}-m\omega_{\rm mag}^2\,\langle X_{\rm CM}^2\rangle/(2\mu_B)$. Using the virial theorem for a harmonic trap, which holds for weakly interacting atoms, we obtain $2 m\,\bar{\omega}_x^2\,\langle X_{\rm CM}^2\rangle=\langle E^x_{\rm CM}\rangle$, where $2m$ is the total mass. Hence,
\begin{equation}
B_{eff}=B_{z0}-\frac{\omega_{\rm mag}^2}{\bar{\omega}_x^2}\frac{\langle E^x_{\rm CM}\rangle}{4\mu_B}-\frac{\langle K^x_{\rm rel}\rangle}{2\mu_B}-\frac{\langle K^{\perp}_{\rm rel}\rangle}{2\mu_B}.
\label{eq:2.5a}
\end{equation}
Here, we have separated the relative kinetic energy term of eq.~\ref{eq:1.5a} into axial and transverse parts.

Next, we evaluate the relative kinetic energy contributions.  For the axial $x$-direction, we {\it select} the energy of the two colliding atoms $E$ and $E'$ in $g(E,E')$, eq.~\ref{eq:15.3}. Hence, the total energy is $E+E'=E^x_{\rm CM}+E^x_{\rm rel}$. For harmonic confinement, the kinetic and potential energies are quadratic degrees of freedom, which requires $E^x_{\rm CM}=(E+E')/2$ for any product state $\phi_E(x_1)\,\phi_{E'}(x_2)$. We also have $E^x_{\rm rel}=(E+E')/2$, where  $E^x_{\rm rel}=K^x_{\rm rel}+\mu\,\bar{\omega}_x^2\,x_{\rm rel}^2/2$ for harmonic confinement. To evaluate $\langle K^x_{\rm rel}\rangle$, we note that for a collision to occur, the relative position $x_{\rm rel}$ of the two atoms must vanish for a contact interaction. Hence, $K^x_{\rm rel}=E^x_{\rm rel}=(E+E')/2$. For the transverse directions,  we have defined a mean fractional spatial density $\bar{n}_\perp$, by eq.~\ref{eq:nperpav}. Assuming that the corresponding relative momentum average for the two transverse directions is determined by a Boltzmann distribution, $\langle K^{\perp}_{\rm rel}\rangle\simeq k_BT$.  Using these results in eq.~\ref{eq:2.5a}, we obtain finally,
\begin{equation}
B_{eff}=B_{z0}-\frac{\langle K^{\perp}_{\rm rel}\rangle}{2\mu_B}-\left(1+\frac{\omega_{\rm mag}^2}{2\bar{\omega}_x^2}\right)\frac{E+E'}{4\mu_B},
\label{eq:2.9a}
\end{equation}
where we leave $\langle K^{\perp}_{\rm rel}\rangle$ as an adjustable parameter, of order $k_BT$.
Replacing $a_S$ with $a(E',E)=a'(B_{eff}-B_0)$ in $g(E',E)$ of eq.~\ref{eq:15.3} and in the results for $\tilde{g}(n',n)$ that follow from it, we obtain a reasonable fit to the high temperature spin density profile of Fig.~6 in the main paper with $\langle K^{\perp}_{\rm rel}\rangle=0.59\,k_BT$.  For $T=45.7\,\mu$K, this corresponds to a shift of $-0.2$ G in $B_{eff}$,  consistent with the upward shift of the applied field for which $a_{12}=0$, as  reported in Table I of the main paper. For the low temperature data, where the energy scale is $<1\,\mu$K,  the corresponding energy shift is negligible.

\subsection{Measured Spatial Profiles versus Predictions}
To compare the data for degenerate samples to the zero temperature theoretical model discussed above, we assume that the measured initial densities $n_\uparrow(x)$, $n_\downarrow(x)$ and the conserved total density are zero temperature Thomas-Fermi profiles (see eq.~\ref{eq:TFdensity}), with an effective zero temperature Fermi radius $\sigma$, which we use as a fit parameter. From the profile of the total density, we find $\sigma=329\,\mu$m, corresponding to an effective Fermi temperature of $m\bar{\omega}_x^2\sigma^2/2=0.82\,\mu$K and a transverse radius $(\bar{\omega_x}/\omega_\perp)\,\sigma$.  For the high temperature sample, the total atom number is $\sim 4.5\times10^5$, and the measured gaussian $1/e$ radius is $\sigma_x=\sqrt{2k_BT/(m\bar{\omega}_x^2)}=325\,\mu$m, which determines $T=45.7\,\mu$K.

\begin{figure*}[htb]
\begin{center}\
\includegraphics[width=6.2in]{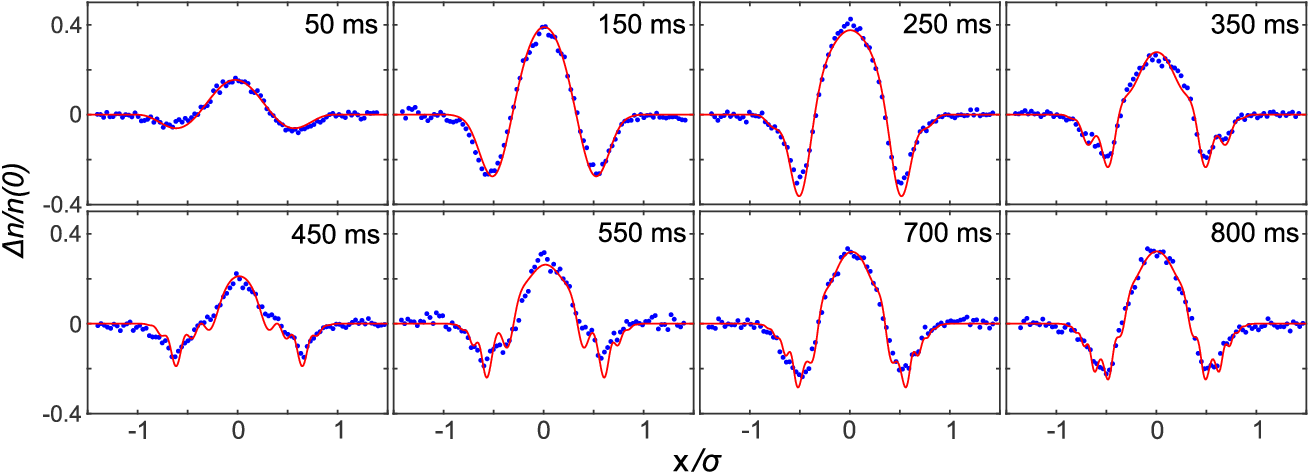}
\end{center}
\caption{Spin density profiles (blue dots) for a degenerate sample $T/T_F=0.35$ versus evolution time relative to coherent excitation. Each data profile is the average of 5 runs, taking in random time order. Each solid red curve is the mean field model with a fixed scattering length of $a=5.23$ bohr ($B=528.844$ G) and a fitted cloud size within a few percent of the average value $\sigma=329\,\mu$m.
\label{fig:TimeDepSpinDensity}}
\end{figure*}

Fig.~\ref{fig:4profilesA} of the main text demonstrates the excellent quantitative agreement between the predicted and measured density profiles of each hyperfine state for a degenerate sample, in units of the conserved average central density $(n_1+n_2)$, for a scattering length of $a_{12}=3.04\,a_0$. Fig.~\ref{fig:SpinDensityProfiles} of the main text shows the transversely integrated  spin densities $n_1(x,t)-n_2(x,t)\equiv 2S_z(x,t)$ with  $a_{12}=5.17\,a_0$ and $a_{12}=-5.39\,a_0$  at selected times $t$ after excitation. The data are quite sensitive to the evolution time and exhibit a complex structure, which are very well fit by the collective spin rotation model.  Fig.~\ref{fig:TimeDepSpinDensity}  shows additional measurements and predictions for the time evolution of $(n_1-n_2)$ between $t=0$ and 800 ms, relative to coherent excitation, for a fixed scattering length of $a=5.23\,a_0$ at $B=528.844$ G, which corresponds to the evolution of the central spin-density shown in Fig.~\ref{fig:amplitude}. Here, $(n_1-n_2)$  is given in units of the total central density $n_1(0)+n_2(0)$.

\end{document}